\documentstyle[amsmath,epsf,psfig,epsfig]{mn}

\newcommand{\beqa}{\begin{eqnarray}}
\newcommand{\eeqa}{\end{eqnarray}}
\newcommand{\f}{\frac}

\def\gsim {\lower .1ex\hbox{\rlap{\raise .6ex\hbox{\hskip .3ex
        {\ifmmode{\scriptscriptstyle >}\else
                {$\scriptscriptstyle >$}\fi}}}
        \kern -.4ex{\ifmmode{\scriptscriptstyle \sim}\else
                {$\scriptscriptstyle\sim$}\fi}}}
\def\lsim {\lower .1ex\hbox{\rlap{\raise .6ex\hbox{\hskip .3ex
        {\ifmmode{\scriptscriptstyle <}\else
                {$\scriptscriptstyle <$}\fi}}}
        \kern -.4ex{\ifmmode{\scriptscriptstyle \sim}\else
                {$\scriptscriptstyle\sim$}\fi}}}
\def\kmsmpc{\,{\rm km\,s^{-1}\,Mpc^{-1}}}
\def\msun{\,{\rm M_\odot}}
\def\lta{\mathrel{\spose{\lower 3pt\hbox{$\mathchar"218$}}
     \raise 2.0pt\hbox{$\mathchar"13C$}}}
\def\gta{\mathrel{\spose{\lower 3pt\hbox{$\mathchar"218$}}
          \raise 2.0pt\hbox{$\mathchar"13E$}}}

\def\etal{et al. }
\def\eg{e.g.}
\def\ie{i.e.}
\def\pa{\partial}
\def\prop{\propto}
\def\apriori{{\it a priori}}

\def\ltsima{$\; \buildrel < \over \sim \;$}
\def\lsim{\lower.5ex\hbox{\ltsima}}
\def\gtsima{$\; \buildrel > \over \sim \;$}
\def\gsim{\lower.5ex\hbox{\gtsima}}
\def\se#1{\S\ref{sec:#1}}
\def\fig#1{Fig.~\ref{fig:#1}}
\def\equ#1{Eq.~(\ref{eq:#1})}
\def\ifm#1{\relax\ifmmode#1\else$\mathsurround=0pt #1$\fi}

\def\hmpc{\,h\ifm{^{-1}}{\rm Mpc}}
\def\hkpc{\,h\ifm{^{-1}}{\rm kpc}}
\def\solar{\ifmmode_{\mathord\odot}\;\else$_{\mathord\odot}\;$\fi}
\def\msun{{\rm M}_{\solar}}
\def\omm{\Omega_{\rm m}}

\def\am{{\bf L}}
\def\bfx{{\bf x}}
\def\bfr{{\bf r}}
\def\bfv{{\bf v}}
\def\bfq{{\bf q}}
\def\bfS{{\bf S}}
\def\bfg{{\bf g}}

\def\D{{\cal D}}

\begin{document}

\title[Testing Tidal-Torque Theory] 
       {Testing Tidal-Torque Theory:
       I. Spin Amplitude and Direction}

\author[C. Porciani, A. Dekel \& Y. Hoffman]
{Cristiano Porciani$^{1,2}$, Avishai Dekel$^1$ and Yehuda Hoffman$^1$\\
$^1$Racah Institute of Physics, The  Hebrew University,
Jerusalem 91904, Israel\\
$^2$Institute of Astronomy, University of Cambridge,
Madingley Road, Cambridge CB3-OHA, UK}

\maketitle
\begin{abstract}

We evaluate the success of linear tidal-torque theory (TTT) in predicting
galactic-halo spin using a cosmological $N$-body simulation with 
thousands of well-resolved haloes.
The proto-haloes are identified by tracing today's haloes back to the 
initial conditions. The TTT predictions for the proto-haloes 
match, on average, 
the spin amplitudes of today's virialized haloes if linear growth is 
assumed until $\sim t_0/3$, or 55-70 per cent of the halo 
effective turn-around time.  This makes it a useful qualitative tool 
for understanding certain average properties of galaxies,
such as total spin and angular-momentum distribution within haloes, 
but with a random scatter of the order of the signal itself. 
Non-linear changes in spin direction cause a mean error of
$\sim 50^\circ$ in the TTT prediction at $t_0$, such that 
the linear spatial correlations of spins on scales $\geq 1\hmpc$ are
significantly weakened by non-linear effects. 
This questions the usefulness of TTT for predicting intrinsic alignments
in the context of gravitational lensing. 
We find that the standard approximations made in TTT, including a second order
expansion of the Zel'dovich potential and a smoothing of the tidal field,
provide close-to-optimal results.

\end{abstract}


\section{Introduction}
\label{sec:intro}

Angular momentum is clearly one of the key physical ingredients
in the process of galaxy formation. The total angular momentum,
and its distribution, must have a crucial role in determining the 
galaxy history and final type. It has therefore been a subject
for classical investigations, 
pioneered by Hoyle (1949),
and then analysed qualitatively by Peebles (1969).
This led to the `standard' theory for the origin of angular 
momentum in the framework
of hierarchical cosmological structure formation, the tidal-torque theory 
(TTT) due to Doroshkevich (1970) and White (1984).  

Special interest in the subject has been revived recently because of a
``spin crisis", arising from cosmological simulations of 
galaxy formation which use hydrodynamical gravitational codes to 
follow the gas dynamics inside dark-matter haloes, and semi-analytical
recipes for star formation and feedback. These simulations seem to yield 
luminous galaxies that are significantly smaller, and of much less 
angular momentum than observed disc galaxies 
(Navarro, Frenk \& White 1995; Navarro \& Steinmetz 1997, 2000).
The discovery of massive black holes in galactic centers 
(Kormendy \& Richstone 1995; Magorrian \etal 1998; Gebhardt \etal 2000a,b) 
provides another motivation for understanding the detailed distribution of
angular momentum within galaxies.
A general need for a detailed recipe for the build-up of angular momentum
in galaxies comes from the developing semi-analytic models,
which have been proven very useful in the study of galaxy formation 
(e.g. White \& Frenk 1991; Kauffmann, White \& Guiderdoni 1993;
Cole et al. 1994; Somerville \& Primack 1999).
Another current motivation comes from weak lensing studies. Our
ability to reconstruct maps of the cosmic mass distribution from
weak gravitational lensing measurements (Bacon, Refregier \& Ellis
2000; Kaiser, Wilson \& Luppino 2000; van Waerbeke \etal 2000, 2001; Wittman
\etal 2000) is hampered by the unknown intrinsic distribution and alignment 
of galaxy shapes.  As a first guess, one crudely assumes that the 
shapes and orientations of background galaxies are uncorrelated in space, 
and that the detected correlations of galaxy ellipticities are solely
due to the lensing by the foreground mass distribution.  However, 
a series of observations (Brown \etal 2000; Pen, Lee \& Seljak 2000)
and numerical simulations (Croft \& Metzler 2000; Heavens, Refregier \& 
Heymans 2000; Dekel \etal 2000)
evidenced that both the intrinsic galaxy shapes and 
their spins are spatially correlated. 
Theoretical models for the alignment of disc galaxies orientations 
tend to assume that they are induced by spin correlations
(Catelan, Kamionkowski \& Blandford 2001; Crittenden \etal 2001).

These issues add a timely aspect to the motivation for revisiting 
the classical problem of angular momentum, in an attempt to sharpen our 
understanding of its various components. This includes the first step
of angular-momentum acquisition by dark haloes, which might be 
particularly assumed to be fairly well understood. 
The basic idea of TTT is that most of the angular momentum is being 
gained gradually by proto-haloes in the linear regime of density fluctuations
growth,
due to tidal torques from neighboring fluctuations, and that this process
continues until the proto-halo reaches its maximum extent. 
The further assumption is that only little angular momentum is being
exchanged between haloes later on in the non-linear regime, after the 
proto-haloes have decoupled from the expanding background and collapsed 
to virialized systems.
It is commonly assumed that the baryonic material, which in general 
follows the dark-matter distribution inside each proto-halo, gains a similar 
specific angular momentum and carries it along when it contracts
to form a luminous galaxy at the halo centre.
This should allow us to predict galactic spins using the approximate
but powerful analytic tools of quasi-linear theory of gravitational 
instability.

In a series of papers,
we evaluate the performance of the TTT approximation, and trace the roles 
of its various ingredients, using a cosmological $N$-body simulation 
with thousands of well-resolved haloes. 
As we do so, we find to our surprise that some of the basic ingredients of TTT,
which are commonly assumed to be 
of `text-book' status,
involve certain unjustified assumptions and confused understandings, 
which may lead to poor approximations. These papers represent attempts to
clarify some of these controversial issues.

We address the TTT at different levels. 
In this paper (Paper I),
we evaluate how well does the approximation predict the final 
angular momentum of a halo, given full knowledge of the corresponding 
initial proto-halo and the cosmological realization.  
In Paper II (Porciani, Dekel \& Hoffman 2001), we attempt 
a deeper level of understanding of the origin of halo angular momentum, 
by investigating the relation between the different components of TTT.
This study connects to the fundamental open question of how to
identify a proto-halo in a given realization of initial conditions.
In Paper III (Porciani \& Dekel in preparation), we revise the standard
scaling relation of TTT, based on the fact that the shear
tensor is only weakly correlated 
with the density contrast of the proto-halo   
(the density and shear being the trace and traceless parts 
of the same, deformation tensor). This scaling relation is used to 
predict the typical angular-momentum profile of haloes (Dekel \etal 2001; 
Bullock \etal 2001b), and to provide a simple way to incorporate angular
momentum in semi-analytic models of galaxy formation (Maller, Dekel \&
Somerville 2002).

Given the initial conditions of a proto-halo and its environment,
TTT predicts the final halo angular momentum based on four specific 
assumptions as follows:
(a)
The flow is laminar, with a one-to-one correspondence of Eulerian
and Lagrangian positions, or the velocity field is properly 
smoothed such that laminarity prevails in practice.
(b) 
The velocities obey the Zel'dovich approximation.
(c) 
The potential at every point within the proto-halo can be approximated by its
Taylor expansion to second order (with respect to spatial separation) 
about the centre of mass.
(d) 
There is little contribution to the halo angular momentum from 
non-linear effects.
In the current paper,
we set to evaluate the global success of TTT in this task,
and try to address the validity of each of the above assumptions.
The role of non-linear effects, in particular, turns out to be the key issue.
This will be an attempt to follow and improve on earlier work 
by White (1984), Hoffman (1986, 1988), Barnes \& Efstathiou (1987), 
Catelan \& Theuns (1996), Lee \& Pen (2000) 
and Sugerman, Summers \& Kamionkowski (2000). 

The outline of this paper is as follows:
In \se{ttt} we summarize the basics of linear tidal-torque theory.
In \se{simu} we describe the simulation and the halo finder.
In \se{implement} we describe the implementation of TTT to proto-haloes.
In \se{amplitude} we evaluate the success of TTT in predicting the amplitudes
  of halo spins.
In \se{direction} we assess the success of TTT in predicting the directions
  of halo spins.
In \se{L-L} we address the spatial coherence of spin directions, which
  is directly relevant to intrinsic alignments in the context of gravitational
    lensing.
In \se{conc} we discuss our results and conclude.

\section{Tidal-Torque Theory}
\label{sec:ttt}

We start by summarizing the basics of linear tidal-torque theory.
The framework is the standard Friedmann-Robertson-Walker cosmological model
in the matter-dominated post-recombination era, in which the matter
is assumed to be a collision-less fluid and the dynamics is driven by 
gravity only. 
In the linear regime, the developing proto-haloes are assumed to be 
small perturbations about the mean universal density.  The peculiar
velocity field, 
due to the growing mode of gravitational instability, describes
a potential flow with no curl, assuming that any primordial source 
of vorticity (\eg, due to some primordial turbulence, see Jones 1976) 
has decayed away due to the expansion of the universe.
 
Given a proto-halo, a patch of matter occupying an Eulerian volume $\gamma$
that is destined to end up in a 
virialized halo, the goal is to compute the halo angular momentum about the 
centre of mass, to the lowest non-vanishing order in perturbation theory.
The angular momentum at time $t$ is defined by
\begin{equation}
\am (t)=\int_{\gamma} \rho(\bfr,t) \left[ \bfr (t)-\bfr_{\rm cm}(t) \right]
\times \left[ \bfv (t)-\bfv_{\rm cm}(t) \right] d^3r \, ,
\label{eq:Lprop}
\end{equation}
where $\bfr$ and $\bfv$ are the position and peculiar velocity vectors,
and the centre-of-mass quantities, 
$\bfr_{\rm cm}$ and $\bfv_{\rm cm}$, 
are defined as usual.
The term proportional to $\bfv_{\rm cm}$ does not contribute to $\am$, 
and it will not be considered any further. 
Re-write \equ{Lprop} in comoving units, 
$\bfx=\bfr /a(t)$ and $\bfv=a\, d \bfx /dt$, 
where $a(t)$ is the universal expansion factor,
and replace $\rho(\bfx,t)$ with the density contrast 
$\delta(\bfx,t) = \rho(\bfx,t) /\bar{\rho}(t) -1$ 
with respect to the average density $\bar{\rho}(t)$, to obtain
\begin{equation}
\am (t) =\bar{\rho}(t) 
a^5(t) \int_{\gamma} \left[ 1+ \delta(\bfx,t) \right]
\left[ \bfx (t)-\bfx_{\rm cm}(t) \right]
\times \dot{\bfx} \, d^3x \, .
\label{eq:Lcomove}
\end{equation}
A dot denotes a derivative with respect to cosmic time, $t$. 
In the matter-dominated era,
$\bar{\rho}(t) a^3(t) = \bar{\rho}_0 a_0^3 = {\rm const}$,
where the subscript 0 denotes quantities evaluated at the present time.
             
In the Lagrangian description of fluid dynamics, the comoving Eulerian 
position of each fluid element 
is given by its initial, Lagrangian position $\bfq$ plus a displacement:
$\bfx(\bfq,t)=\bfq+\bfS(\bfq,t)$.
When fluctuations are small, or when the flow is properly 
smoothed, 
the mapping $\bfq \to \bfx$ is reversible and the flow is {\it laminar}. 
Then the Jacobian determinant $J=|| \pa \bfx/\pa \bfq ||$ does 
not vanish, and the continuity equation implies
$1+\delta[\bfx(\bfq,t)]=J^{-1}(\bfq,t)$.
Substituting in \equ{Lprop} we obtain, for a laminar flow,
\begin{equation}
\am (t)=a^2(t)\,\bar{\rho}_0 a^3_0 \int_{\Gamma}
\left[\bfq-\bar{\bfq}+\bfS(\bfq,t)-\bar{\bfS}\right]
\times \dot{\bfS}(\bfq,t) \,d^3q\, ,
\label{eq:Llag}
\end{equation}
where $\Gamma$ is the region in Lagrangian space corresponding to $\gamma$ in
Eulerian space.  Barred quantities are averages over $\bfq$ in $\Gamma$.
Note that \equ{Llag} is exact in the absence of orbit crossing.

The displacement $\bfS$ is now spelled out using the {\it Zel'dovich\,} 
approximation (Zel'dovich 1970), which is linear in Lagrangian terms 
and mildly non-linear in Eulerian space.  
It assumes that the proportionality between the
velocity potential and the gravitational potential $\phi(\bfq,t)$, 
which holds in the linear regime for the growing mode of density fluctuations 
$\delta \prop D(t)$, can be extended into the mildly non-linear regime.
This implies that the spatial and temporal dependences of $\bfS$ can be 
decoupled,
\begin{equation}
\bfS(\bfq,t)=-D(t)\, {\bf \nabla} \Phi(\bfq) \, .
\label{eq:ZA}
\end{equation}
where $\Phi(\bfq)= \phi(\bfq,t)/ 4 \pi G \rho(t) a^2(t) D(t)$
(with $G$ Newton's gravitational constant). 
It also implies that the peculiar velocity and acceleration fields 
are parallel.
Substituting \equ{ZA} into \equ{Llag} one obtains  
\begin{equation}
\am (t)=-a^2(t) \dot{D}(t)\,\bar{\rho}_0 a^3_0 \,
\int_{\Gamma} (\bfq-\bar{\bfq}) \times {\bf \nabla} \Phi(\bfq) \,d^3q\, .
\label{eq:LZel1}
\end{equation}
We see that the explicit growth rate is $L \propto a^2(t) \dot{D}(t)$. 
For an Einstein-de Sitter universe this is $\prop t$.\footnote{
A numerical integration of the cosmological relations shows that for all flat
cosmologies the approximation $a^2 \dot D \prop D^{3/2}$ holds to an
excellent accuracy, and it deviates only slightly for open cosmologies.}

Next, assume that the potential is smoothly varying within the volume
$\Gamma$, such that it can be approximated by its {\it second order\,} Taylor 
expansion about the centre of mass $\bar{\bfq}$,
\begin{equation}
\Phi(\bfq') \simeq \Phi({\bf 0})+\left.\frac{\pa \Phi} {\pa q'_i}
\right| _{\bfq' =0} \,q'_i+
\frac{1}{2} \left.
\frac{\pa^2 \Phi}{\pa q'_i \pa q'_j} \right| _{\bfq'=0}
\,q'_i \, q'_j\;, 
\label{eq:phiexp}
\end{equation}
where 
$\bfq'\equiv \bfq-\bar{\bfq}$.\footnote{
This is equivalent to assuming that the velocity field is well described
by its linear Taylor expansion, i.e., 
$v_i-\bar{v}_i\simeq\D_{ij}\,q'_j$
(where $v$ is in comoving units). 
}
Substituting in \equ{LZel1} one obtains the basic TTT expression
for the $i$th Cartesian component:
\begin{equation}
L_i (t)= a^2(t) \dot{D}(t)\, \epsilon_{ijk}\, \D_{jl}\, I_{lk} \, ,
\label{eq:TTT}
\end{equation}
where $\epsilon_{ijk}$ is the fully antisymmetric rank-three tensor,
and the two key quantities are the {\it deformation\,} tensor at 
$\bfq'=0$, 
\begin{equation}
\D_{ij}=-\left.\frac{\pa^2 \Phi}{\pa q'_i \pa q'_j} 
\right| _{\bfq'=0}\,, 
\label{eq:deformation}
\end{equation}
and the {\it inertia\,} tensor of $\Gamma$,
\begin{equation}
I_{ij}=\bar{\rho}_0 a^3_0 
\int_{\Gamma} q'_i \,q'_j \,d^3q' \, . 
\label{eq:inertia}
\end{equation}
Note that only the traceless parts of the two tensors matter for the
cross product in \equ{TTT}. These are the velocity {\it shear\,}
or {\it tidal\,} tensor, $T_{ij}= \D_{ij}-(\D_{ii}/3) \delta_{ij}$,
and the traceless quadrupolar inertia tensor, $I_{ij}-(I_{ii}/3)\delta_{ij}$.
Thus, to the first non-vanishing order in perturbation theory,
angular momentum is transferred to the developing proto-halo by the
gravitational coupling of the quadrupole moment of its mass distribution
with the tidal field exerted by neighboring density fluctuations. 
The torque depends on the proto-halo shape, the external tidal field,
and the {\it misalignment\,} between the two.
In particular, $\am$ vanishes if $\Gamma$ is a sphere or is bounded by an
equipotential surface of $\Phi$. 

There is no \apriori\, justification for truncating the 
expansion of the potential after second-order, \equ{phiexp}.
More generally, 
the expansion includes further terms,
\begin{eqnarray}
L_i (t)&=&
L_i^{(2)}(t)+L_i^{(3)}(t)+\dots= \\ 
&=& a^2(t) \dot{D}(t)\, 
\epsilon_{ijk} \left[ \D_{jl} I_{lk} + \frac{1}{2}
{\cal D}^{(3)}_{jlm}I^{(3)}_{lmk}+\dots \right]\,, \nonumber
\label{eq:LZel3}
\end{eqnarray}
with 
\begin{equation}
D^{(3)}_{ijk}= -\left.{\pa^3 \Phi \over \pa q'_i \pa q'_j \pa 
q'_k} 
\right| _{\bfq'=0} 
\!\!\!\!,\quad 
I^{(3)}_{ijk}= \bar{\rho}_0 a^3_0 \int_{\Gamma} q'_i q'_j 
q'_k\,d^3q' 
\,.
\end{equation}
Here, multipole moments of a given order of the proto-halo mass distribution 
are coupled with spatial derivatives of the same order of the potential.
This is a consequence of the $1/r^2$ scaling of gravitational interactions, 
and indeed, the total torque for a localized charge in an external 
electrostatic field admits a similar expansion (\eg, Jackson 1975,
exercise 4.5).
Note that dipole interactions [\ie, terms proportional to
${\bf p} \times \bfg(\bfq'=0)$ 
with ${\bf p}$ the mass dipole moment and
$\bfg$ the gravitational acceleration] do not contribute to the torque 
because the spin is considered in the centre-of-mass frame, where the mass
dipole vanishes. On the other hand, dipole interactions are expected to play
an important role in the evolution of orbital angular momentum.

After the dark-matter haloes have detached from the expanding background
and collapsed into condensed systems, the tidal torquing continues, but
at a highly reduced rate. 
Peebles (1969) estimated the typical torque exerted on a quadrupolar mass 
distribution by a population of point-like galaxies, showing that the 
transfer of angular momentum becomes less and less efficient with time.
This is a consequence of the fact that haloes become small when they collapse,
as well as the fact that they tend to move away from each other
due to the Hubble expansion.

\section{Simulation and Halo Finding}
\label{sec:simu}

The $N$-body simulation analysed here was performed as part of the
GIF project (\eg, Kauffmann \etal 1999) using the parallel, adaptive
particle-particle/particle-mesh (AP$^3$M) code developed by the Virgo
consortium (Pearce \etal 1995; Pearce \& Couchman 1997).
As an example, we use a simulation of the $\tau$CDM scenario,
in which the cosmology is flat, Einstein-de Sitter with density parameter
$\omm=1$, and Hubble constant $H_0=100 \,h \kmsmpc$ with $h=0.5$.
The power spectrum of initial density fluctuations is CDM with shape 
parameter $\Gamma=0.21$.
The amplitude of the power spectrum is set such that the rms overdensity
in top-hat spheres of radius $8\hmpc$, as extrapolated using linear theory
to the present epoch, is $\sigma_8=0.51$.\footnote{
Such a power spectrum may arise, for example, if a late decay of massive 
$\tau$ neutrinos produces an additional background of relativistic $e$ and 
$\mu$ neutrinos, thus delaying the onset of matter domination and lowering
the effective value of $\Gamma$ (Bond \& Efstathiou 1991; 
Efstathiou, Bond \& White 1992; White, Gelmini \& Silk 1995).
Here it simply serves us as a scenario which obeys many of the observational
constraints from large-scale structure, including the normalizations set by
cluster abundance and by COBE's measurements of large-angle anisotropies 
in the CMB, independently of its physical origin.}

The simulation was performed in a periodic cubic box of side $84.55\hmpc$,
with the mass represented by $256^3$ particles of 
$1.0 \times 10^{10} h^{-1} M_\odot$ each.
Long-range gravitational forces were computed on a $512^3$ mesh, while
short-range interactions were calculated as in Efstathiou \& Eastwood (1981).
At late times ($z \lsim 3$), the corresponding gravitational potential 
asymptotically matches a Plummer law with softening $\epsilon=36\hkpc$
both at large and small scales. The simulation started at a redshift $z=50$ 
and ended at the present time, $z=0$.
The initial conditions were generated by displacing
particles according to the Zel'dovich approximation from an initial 
stable `glass' state (e.g. White 1996).
More details are given in Jenkins \etal (1998).

We identify virialized dark-matter haloes at $z=0$ as follows.
First, we select an initial set of groups of particles
using a friends-of-friends method (\eg, Davis \etal 1985)
with a linking length $b=0.2$ in units of the average inter-particle distance.
This algorithm identifies regions bounded by a surface
of approximately constant density contrast, 
$\rho/\bar\rho\simeq 3/(2\pi b^3)$ (e.g. Cole \& Lacey 1996).  
Assuming that the density profiles of these groups
approximate 
singular isothermal spheres, the selection criterion can be expressed in terms
of the average density contrast in the halo, 
$\rho/\bar\rho \simeq 9/(2\pi b^3)$.
Thus, $b=0.2$ corresponds to haloes with a mean density contrast 
$\simeq 179$, in general 
agreement with the predictions of the spherical collapse model 
for perturbations whose outer shells have collapsed recently.

We then remove unbound particles from each halo by the following iterative
procedure. At each step, we compute the energies of all particles in the
current centre-of-mass frame, and remove the most unbound particle.
We then re-compute the energies of all the particles and so on.
The iteration stops when no unbound particles are found.
Note that a particle that has been removed is allowed
to re-enter the halo if it becomes bound at a later step
due to changes in the position and velocity of the centre of mass.
Clumps which end up containing between 100 and 200 particles loose, 
on average, $\sim 5.5$ per cent of their mass as unbound particles.
In a few cases, typical of small haloes, the friends-of-friends 
algorithm identifies `fake' haloes which are not bound gravitationally. 
For example, seven friends-of-friends haloes (out of $\sim 7300$) 
composed by more than 100 particles end up 
containing only 10 bound particles or less
after the un-binding procedure.
More massive haloes are, naturally, less affected by unbound interlopers
because their potential wells are typically deeper.
For instance, the mean fraction of unbound mass removed from clumps which
end up containing between 200 and 1000 (1000-3000) particles is $\sim 2.9$
per cent ($\sim 2.0$ per cent).
In the following we consider only haloes which contain more than 100
particles after the un-binding procedure is applied.
Our conclusions regarding TTT turn out to be insensitive to the removal
of unbound particles, and to increasing the minimum-mass threshold.
In order to try to restrict ourselves to galactic haloes, we also exclude 
all the $\sim 200$ haloes which consist of more than 3000 particles --
these tend to have more substructure and are likely to correspond to groups 
or clusters of galaxies. 

The robustness of our conclusions regarding TTT with respect to the 
halo-finding algorithm (\eg, friends-of-friends versus fitting a spherical
or ellipsoidal density profile, Bullock \etal 2001a) is investigated
in another paper in preparation.
Note that all the haloes in our current sample are not subclumps of larger 
host haloes, for which linear theory is not expected to be valid.
This excludes about $10$ per cent of the haloes that are more massive
than $10^{12}\, h^{-1} \msun$ (Sigad \etal 2001).

\section{Implementing TTT}         
\label{sec:implement}

The proto-halo regions $\Gamma$ in Lagrangian space are defined 
straightforwardly by tracing all the halo particles into their 
Lagrangian positions.  
Generally, most of the 
proto-halo mass is contained
in a simply connected Lagrangian region, 
but in some cases the proto-halo may be 
divided into several 
compact regions which are connected by thin filaments. 
Nearly 10 per cent of the proto-haloes are characterized by extended
filaments departing from a compact core. 

We stress that the proto-haloes at earlier times are defined solely based
on their identification as virialized haloes at $z=0$, without worrying
about the details of the merger history, and without applying any
other criteria at the initial conditions.
Some of the results reported below are specific to this choice of
considering all today's halo particles at all times rather than, e.g., 
investigating the history of the halo major progenitor (Vivitska et al. 2001; 
Wechsler et al., in preparation).

For each proto-halo, we compute the Lagrangian inertia tensor 
by direct summation over its $N$ particles of mass $m$ each:
\begin{equation}
I_{ij}=m \sum_{n=1}^{N} q_i^{'(n)} q_j^{'(n)}\;, 
\end{equation}
with $\bfq^{'(n)}$ the position of the $n$-th particle with respect
to the halo centre of mass.

We use three alternative methods to measure the shear tensor
at the proto-halo centre of mass in the initial conditions of the 
$N$-body simulation.

{\it Method 1, top-hat smoothing:} 
We smooth the gravitational potential
used to generate the initial Zel'dovich displacements, and then
differentiate it two times with respect to the spatial coordinates
to obtain the deformation tensor.
Smoothing is done using a top-hat window function, while derivatives are
computed on a grid. The top-hat smoothing radius in comoving units, 
$R_{\rm th}$,
is taken to be defined by $(4 \pi/3) \bar\rho_{0} R_{\rm th}^3=M$, 
with $M$ the virial halo mass.
This is the only practical method for application in analytic and semi-analytic
modelling of galaxy formation. We refer to is as the {\it standard\,}
method of TTT.

{\it Method 2, fit total:} 
The smoothing applied in Method 1 is associated with a loss of information.
First, the coarse-graining procedure averages over small-scale structure. 
Second, the spherical smoothing kernel does not maintain
the details of the proto-halo shape, since, as we find in Paper II, 
most proto-haloes are far from being spherically symmetric.
We minimize smoothing in the computation of the tidal tensor in the 
following way.
The comoving linear velocity field, in the proto-halo centre-of-mass frame,
is expanded in terms of the corresponding Eulerian comoving separation,
$\bfx$, as
\begin{equation}
v_i={\cal D}_{ij} \,x_j\;.
\end{equation}
The components of the deformation tensor
${\cal D}_{ij} = \pa v_i / \pa x_j $
are computed by least-squares fitting to the simulation data (cf. Kaiser
1991). 
This is realized by solving the system of equations
\begin{eqnarray}
\label{eq:least}
\sum_{n=1}^{N}
{\cal D}_{ik} x_k^{(n)} x_i^{(n)} &=& \sum_{n=1}^{N} v_i^{(n)} x_i^{(n)}\;, 
\nonumber \\ 
\sum_{n=1}^{N}[ {\cal D}_{ik} x_k^{(n)} x_j^{(n)} +{\cal D}_{jk} x_k^{(n)} x_i^{(n)}]
&=& \\
\sum_{n=1}^{N} [v_i^{(n)} x_j^{(n)}&+& v_j^{(n)} x_i^{(n)}]\;. \nonumber
\end{eqnarray}
Since the linear growing mode of the gravitational velocity field 
is curl-free, \equ{least} has been derived after imposing
the condition ${\cal D}_{ij}={\cal D}_{ji}$.
We then compute the symmetric, traceless shear tensor
$T_{ij} \equiv {\cal D}_{ij}-(1/3) {\cal D}_{ii}\delta_{ij}$.
The shear field computed this way corresponds to minimal smoothing,
at the level of the individual particles in the $N$-body simulation.
However, even though no velocity averaging is performed,
method 2 is, in some sense, similar to using a smoothing kernel
which adapts itself to the proto-halo shape.
The resulting $T_{ij}$ describes the quadrupolar structure
of the velocity field in the proto-halo at the particle level.
Method 2 is therefore expected to provide better predictions for $\am$ 
than Method 1 whenever the top-hat smoothing is associated with
a severe loss of information.

\begin{figure}
\epsfxsize= 8 cm \epsfbox{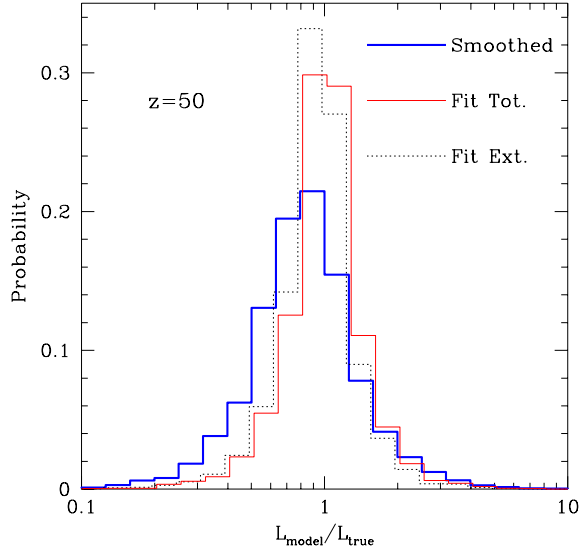}
\caption{
Predicted versus actual angular-momentum amplitude at the initial time, 
$z=50$.
Shown is the distribution over all the haloes in our sample.
The three histograms correspond to the different methods of computing
the shear tensor discussed in the main text (\se{implement}): 
using top-hat smoothing (thick line)  or fitting with no
smoothing of either the total shear (thin line) or the external part only
(dotted line). 
To improve readability, the thin and dotted histograms have been slightly 
displaced in the horizontal direction.
}
\label{fig:ampz50}
\end{figure}

{\it Method 3, fit external:} 
The interactions between particles that belong to the same proto-halo
cannot create any net angular momentum.  This motivates us to 
consider the shear tensor generated solely by the density perturbations 
lying outside the proto-halo volume, $\Gamma$.
This is realized by decomposing the linear velocity field into two parts,
$\bfv=\bfv^{\rm (self)}+\bfv^{\rm (ext)}$, accounting for the 
gravitational pull of the mass lying in and outside the proto-halo, 
respectively.
In practice,
for each particle in $\Gamma$, we first compute the linear gravitational 
acceleration generated at $z=50$
by all the other point masses forming the proto-halo.
>From this quantity, we then subtract the acceleration caused by the same 
particles as if they were part of a uniform background. This is computed
using the unperturbed particle positions of the glass state which,
formally, correspond to $z\to \infty$.
Rescaling the result we obtain $\bfv^{\rm (self)}$.
The difference between the actual velocity of the particles in
the simulation at $z=50$ and this ``self" gravitationally induced term 
is taken to represent $\bfv^{\rm (ext)}$.
Indeed, our estimate for $\bfv^{\rm (ext)}$ generates, in practice, 
the same spin as the total velocity field.
Eventually, the external tidal field is computed by least-squares fitting.
Note that, since ${\cal D}_{ij}^{\rm (self)}\propto I_{ij}$, 
the self field does not contribute to the angular momentum also
at a linear level -- i.e. there is no self-torquing. 

The results obtained by Method 2 and Method 3 
contain spurious contributions due to errors in the 
determination of the deformation
tensor by the fit of a linear model to a discrete set of points. 
However, since the external shear is expected to vary
more smoothly than the self gravity on a proto-halo scale, 
${\cal D}_{ij}^{\rm (self)}$ should be described more accurately by the 
linear expansion of the velocity field. 
For this reason, Method 3 should provide a better estimate of the
angular momentum.
Note, however, that additional noise is introduced in Method 3 by
replacing the uniform background density distribution
with the glass state.

\section{Spin Amplitude}
\label{sec:amplitude}

\def\ltrue{L_{\rm true}}
\def\lmod{L_{\rm model}}

Given a proto-halo at the initial conditions, we first investigate 
the performance of TTT, \equ{TTT}, 
in terms of predicting the amplitude of the halo angular momentum.
Any systematic deviation of the mean predicted amplitude from 
that obtained in the simulations can be eliminated by adjusting
the effective time near turn-around at which the linear TTT growth 
is assumed to stop. Random deviations, both due to the approximations
made already at the initial conditions as well as non-linear effects
at late epochs, cannot be avoided and should be estimated.

We start by testing the accuracy of TTT in the linear regime,
at the {\it initial\,} time, $z=50$. 
For each halo, we compute the inertia tensor $I_{ij}$ and the shear 
tensor $T_{ij}$ using the three different methods described in \se{implement}.
We then compute the amplitude $\lmod$ as predicted 
by \equ{TTT}, and compare it with the actual angular momentum of the 
proto-halo particles, $\ltrue$. The distribution of the ratio of
the two over the protohaloes is shown in \fig{ampz50}, and characterized
in Table~\ref{Apct}.
The mean (median) values for methods 1, 2, 3 are
0.95 (0.83), 1.05 (0.99), 1.01 (0.95) respectively. 
In all cases the probability distribution is positively skewed,
and is leptokurtic 
(with a sharp peak and long tails)
with respect to a Gaussian distribution in log-space. 
It is therefore not very useful to quantify the dispersion about 
the mean by a single standard deviation, so we quote instead 
the 68.3 per cent confidence interval, obtained by locating 
the values corresponding to the 15.9 and 84.2 percentiles in the
cumulative distribution.  For methods 1, 2, 3 we find 
$(0.53,1.28)$, $(0.72,1.31)$, $(0.71,1.24)$ respectively. 
As a single measure of scatter we use half the 68.3 per cent range,
which is $0.37$, $0.29$, $0.26$ respectively. 
All three methods involve the second-order truncation of the expansion
of the potential and some smoothing. These are the only sources of error,
since the Zel'dovich approximation is exact at the initial conditions
of the simulation by construction.
Even though the cases with minimum smoothing (methods 2 and 3) 
involve noisy fits of the small-scale structure, they give somewhat 
more accurate predictions.  The external fit does just 
a little better than the total fit, both in terms of systematic error 
and scatter.  However, the smoothed tidal field (method 1), which is 
the only practical method for semi-analytic modelling of galaxy formation, 
also leads to a small systematic deviation of the mean, of only $\sim 5$ 
per cent.  The scatter, though, is of the order of the signal for all 
three methods already at the initial conditions.

These numbers change only slightly when restricting the analysis
to the very well-resolved haloes containing 1000 particles or more.
In this case, the mean (median) of the distribution of the amplitude ratio is 
$0.90$ (0.79), $1.06$ (1.02), $1.02$ (0.98)   
for methods 1, 2, 3 respectively.
The corresponding scatter is $0.32$, $0.30$, $0.27$.  

When we add the {\it third-order term} in the Taylor expansion of the 
gravitational potential, i.e., the second term in the right-hand side
of \equ{LZel3}, we find no significant improvement in the systematic 
error and in the scatter. In this case the mean (median) and scatter
of $\lmod/\ltrue$ are 0.95 (0.83) $\pm 0.36$, almost identical to the
second-order results (top-hat smoothing is adopted in both cases).
A more detailed inspection reveals the following.
When considering only those haloes for which standard second-order TTT 
over-predicts or under-predicts the 
spin amplitude by more than a factor of 3, we find that by adding the 
third-order term one improves the prediction in about 75 per cent 
of the cases, though the overall scatter increases.
In the case of severe over-prediction, the mean (median) and scatter 
are 4.54 (3.41) $\pm 0.44$ for second-order TTT
and 3.81 (2.63) $\pm 0.79$ for third-order TTT.
In the case of severe under-prediction, the numbers are
0.25 (0.26) $\pm 0.07$ for second-order TTT and 
0.40 (0.33) $\pm 0.16$ for third-order TTT.
On the other hand, when the second-order TTT predictions are reasonably
good, the addition of the third-order term makes no general improvement.
For the haloes were the prediction is within 30 per cent of the true value,
the mean (median) and scatter are
0.94 (0.92) $\pm 0.17$ for second-order TTT and
0.94 (0.91) $\pm 0.21$ for third-order TTT.
Part of the failure of the third order term to improve the results
in these cases could be attributed to numerical errors in the computation 
of the third-order term, which are particularly important when the high-order 
corrections are small.  

\begin{figure}
\epsfxsize= 8 cm \epsfbox{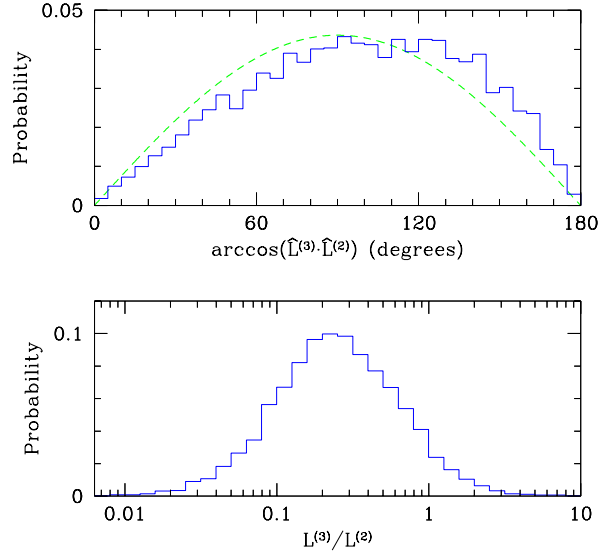}
\caption{Probability distributions for the angle (top)
and the ratio of the amplitudes (bottom)
between the halo spin predicted by TTT considering the first 
term on the right hand side of \equ{LZel3}, 
$\am ^{(2)}$, 
and the second term, 
$\am^{(3)}$.   
The dashed line in the top panel is the expected distribution for random
orientations. 
}
\label{fig:tay3}
\end{figure}

\fig{tay3} shows the distribution of the angle between 
the direction of $\am$ as predicted by standard second-order TTT (top-hat
smoothing) and by the third-order term alone. 
We see that the additional contribution points at an almost random direction
compared to the leading-order term, with an average angle of 
$98^\circ$ and a scatter of $\sim 45^\circ$. 
\fig{tay3} shows the distribution of the corresponding amplitude ratio, 
$L^{(3)}/L^{(2)}$,  
with an average of $0.39$ and a scatter of $0.27$. 
These indicate that any truncation of the expansion is doomed to introduce
an error that is not negligible compared to the signal.
We conclude that the second-order expansion is 
the most sensible and practical procedure, but it is 
associated with a noticable error.  

\begin{figure}
\epsfxsize= 8 cm \epsfbox{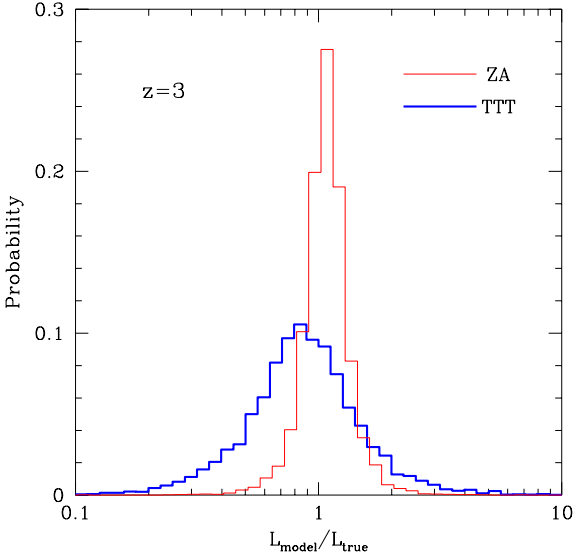}
\caption{ 
Model versus true 
angular-momentum amplitudes in the quasi-linear regime, $z=3$.
Shown is the distribution over all the haloes in our sample.
The thick histogram corresponds to standard TTT based on the
truncated and smoothed shear tensor.
The thin histogram (which has been slightly shifted to the right)
corresponds to $L$ computed directly from the $z=3$ Zel'dovich velocities.
The deviations from unity in the latter is solely due to the non-linear 
effects in the growth rate of angular momentum.
}
\label{fig:ampz3}
\end{figure}

In the second step of our testing,
we address the growth rate indicated by TTT in the {\it quasi-linear\,} regime, 
$L \propto a^2 \dot D$. 
We do it by comparing for each proto-halo the 
model and true 
$L$ values at $z=3$, where the fluctuations are still linear on the Lagrangian
scale of the protohaloes, but the proto-halo substructures, and maybe even
the haloes themselves, already involve non-linear fluctuations.
We show in \fig{ampz3} the distribution of the ratio $\lmod$
to $\ltrue$ for standard TTT.
We show, for comparison, the corresponding
distribution for the case where the model angular momentum is computed directly 
from the $z=3$ velocities, as predicted by the Zel'dovich approximation
from the initial conditions with the corresponding growth. 
 
The deviation of the ZA histogram from a Dirac $\delta$ function at unity
reflects the deviation of the angular-momentum growth rate from the linear
prediction, $L \propto a^2 \dot D$. 
It does not involve the errors associated with the smoothing
and the truncation of the potential because the initial conditions also
obey the Zel'dovich approximation. 
The predicted growth rate is slightly higher than the true growth rate,
but the corresponding 
$8$ per cent systematic overestimate and the $0.19$ scatter 
are relatively small.
The additional effects of truncation plus smoothing in TTT 
are responsible for the larger scatter in the TTT histogram. 
In this case, the systematic error is only 2 per cent, due
to a slight systematic underestimate by TTT of the spin amplitude at $z=50$,
which partly compensates for the overestimate of the growth rate.
Note, however, that the median of the distribution (see Table \ref{Apct}) keeps
roughly constant between $z=50$ and $z=3$. 
The scatter of $0.43$ is comparable to the signal, and is slightly
larger than the initial scatter of $0.38$ due to the deviations from    
the linear growth rate.

\begin{table*}
\begin{minipage}{123mm}
\caption{Characteristics of the distribution of the ratio of amplitudes 
of the spins as 
predicted by the model and as measured in the simulations.
Listed are the mean and several percentiles, as well as the scatter
as measured by half the central 68.3 per cent region.
TTT=tidal-torque theory, NB=$N$-body simulations, ZA=Zel'dovich 
approximation.} 
\label{Apct}
\begin{tabular}
    {@{}lccccccccc}
Comparison  & $z$ & Fig. & {\bf Mean}                
        & $2.5\%$ & $15.9\%$ & ${\mathbf 50\%}$   & $84.2\%$  & $97.5\%$ & 
{\bf Scatter}
\\
\\
TTT/NB  & 50 & 1 &{\bf 0.95}  &0.27 & 0.53 & {\bf 0.83} & 1.28 & 2.48 & 
{\bf 0.37}\\
TTT(2)/NB  & 50 & 1 &{\bf 1.05} &0.43 & 0.72 & {\bf 0.99} & 1.31 & 2.12 & {\bf 0.29}\\
TTT(3)/NB  & 50 & 1 &{\bf 1.01} &0.41 & 0.71 & {\bf 0.95} & 1.24 & 1.98 & {\bf 0.26} \\ 
\\
ZA/NB & 3 & 3 &{\bf 1.08} & 0.65 &0.87 & {\bf 1.05} & 1.25 & 1.67 & {\bf 0.19} \\
TTT/NB & 3 & 3 &{\bf 1.02} & 0.27 &0.53 & {\bf 0.86} & 1.39 & 2.73 & {\bf 0.43} \\
\\
NB/NB $\times 100$& 50/0& - &{\bf 1.17} & 0.26 & 0.52 & {\bf 0.96} & 1.72 & 3.33 &{\bf 0.60}\\
NB/NB& 3/0& - & {\bf 0.50} & 0.13 & 0.23 & {\bf 0.41} & 0.72 & 1.41 & {\bf 0.24}\\
NB/NB& 1.2/0& - &{\bf 1.05} & 0.32 & 0.55 & {\bf 0.90} & 1.45 & 2.76 & {\bf 0.45}\\
NB/NB& 0.6/0& - & {\bf 1.19} & 0.46 & 0.75 & {\bf 1.08} & 1.53 & 2.72 & {\bf 0.39} \\
NB/NB& 0.2/0& - &{\bf 1.08} & 0.65 & 0.90 & {\bf 1.04} & 1.23  & 1.70& {\bf 0.16}\\ 
NB/NB& 0/3& 4 & {\bf 2.93} & 0.70 & 1.38 & {\bf 2.41} & 4.26 & 7.89 & 
{\bf 1.44}\\
\\
ZA/NB &0& 6a & {\bf 1.68} & 0.28 & 0.61 & {\bf 1.20} & 2.39 & 5.44 & {\bf 0.89}
\\
ZA/NB & 0& 6b & {\bf 1.64} & 0.32 & 0.66 &{\bf  1.23} & 2.39 & 5.17 & {\bf 0.86}
\\
ZA/NB & 0& 6c & {\bf 1.47} & 0.33 & 0.66 & {\bf 1.22}  & 2.16 & 4.19 &{\bf 0.75}
\\
\\
TTT/NB & 0 & 6a & {\bf 1.41} & 0.21 & 0.48 & {\bf 1.00} & 2.04 & 4.91 &
{\bf 0.78}\\
TTT/NB & 0 & 6b & {\bf 1.38} & 0.24 & 0.52 & {\bf 1.00} & 2.03 & 4.75 & 
{\bf 0.76}\\
TTT/NB &  0 & 6c &{\bf 1.28} & 0.23 &0.51 & {\bf 1.00} & 1.92& 4.18 &
{\bf  0.70} 
\end{tabular}
\end{minipage}
\end{table*}

In the third step, we compare the spin amplitudes at $z=0$,
after non-linear processes have affected the true spin.
We first wish to determine the effective time in each halo history, 
or on average for all haloes, at which the linear growth of angular 
momentum \`a la TTT should be assumed to stop 
in order to obtain the best prediction for its amplitude at $z=0$.
Then, we wish to evaluate the scatter about the mean TTT result.

\begin{figure}
\epsfxsize= 8 cm \epsfbox{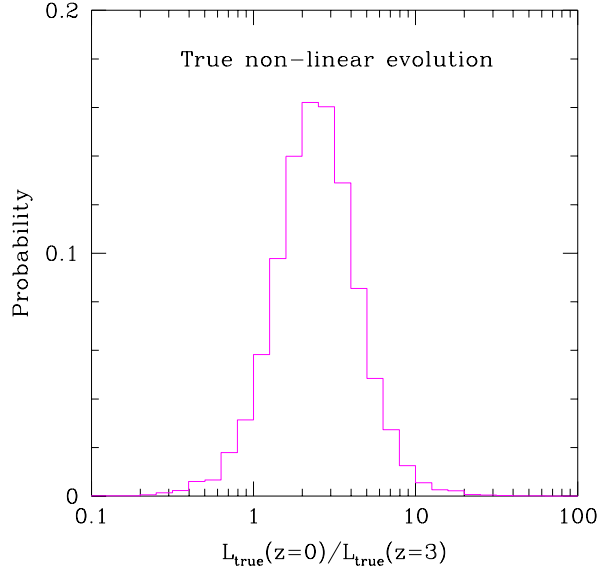}
\caption{
True growth factor of angular momentum between $z=3$ and $z=0$.
Shown is the distribution over all the haloes in our sample.
}
\label{fig:ampz_0_3}
\end{figure}

In order to learn about the actual evolution in the non-linear regime,
we show in \fig{ampz_0_3} the distribution over the haloes
of the ratio between $\ltrue$ at $z=0$ ($t=t_0$) and at $z=3$ ($t=t_0/8$). 
The actual growth factor between these times has a mean value of 2.9 
(with a scatter of $\sim 1.4$), which implies that the linear growth 
should be stopped at an effective time $2.9 \,t_0/8 \simeq 0.36 \,t_0$ 
(or $z\simeq 1$).  This time is somewhat earlier than maximum expansion 
for the outer shell of the halo (using the terminology of the spherical 
collapse model), because our haloes, by definition, have their outermost 
shells reaching virialization today, and were therefore at maximum expansion
at $t \simeq t_0/2$ (or $z \simeq 0.6$). Linear growth until maximum expansion
would have implied a growth factor of 4 instead of 2.9 since $z=3$.
 
\begin{figure} 
\epsfxsize= 8 cm \epsfbox{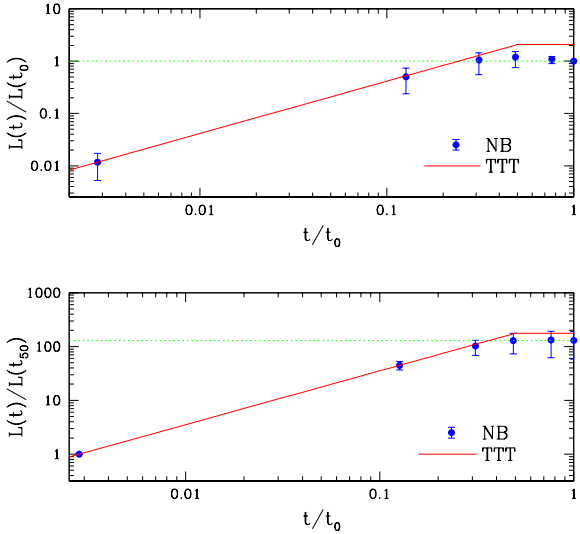}
\caption{
True evolution of spin amplitude in the simulation compared to TTT.
Shown are the average and the 68.3 per cent confidence interval 
over the haloes at different times, of the ratio of spin amplitude $L(t)$ 
and either the spin today (top panel) or the spin at the initial conditions
(bottom panel). The continuous lines represent the linear
TTT growth stopped at maximum expansion ($t=0.5 \,t_0$).
The dotted lines mark the average value at $t_0$.
}
\label{fig:meanevo_amp}
\end{figure}

The actual evolution of spin amplitudes of the simulated haloes is shown in
\fig{meanevo_amp}.
The average and scatter refer alternatively to the ratio
of $L(t)$ and the initial proto-halo spin, and the ratio of $L(t)$ and
the final halo spin. We see that, in agreement with the results shown 
in \fig{ampz_0_3}, the best effective time for stopping 
the TTT growth is $\sim 0.3 \,t_0$. This would properly adjust the mean,
but would leave a significant scatter from halo to halo as indicated by
the error bars.

\begin{figure}
\epsfxsize= 8 cm \epsfbox{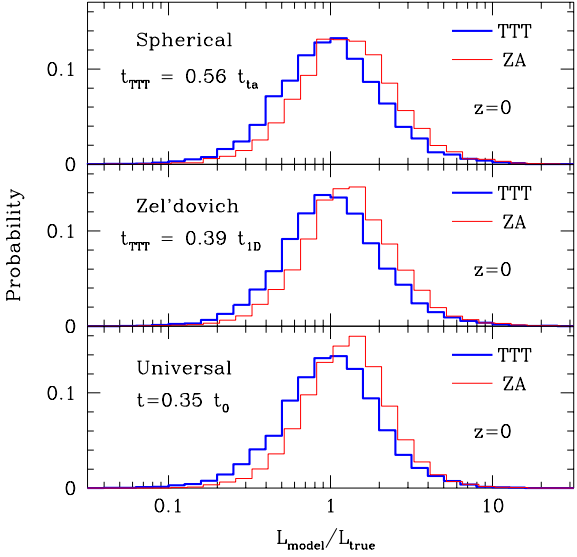}
\caption{ 
Model versus true spin amplitudes in the non-linear regime, at $z=0$.
The TTT growth has been stopped at $t_{\rm TTT}$ such that the median
of the ratio $\lmod/\ltrue$ is unity.
In the top panel   $t_{\rm TTT} = 0.56\, t_{\rm ta}$, a fixed fraction of 
the turn-around time as predicted for each halo by the spherical-collapse 
model based on the smoothed initial density fluctuation. 
In the middle panel $t_{\rm TTT}=0.39\, t_{\rm 1D}$, a fixed fraction of
the time of orbit crossing along the first 
principal axis of the deformation tensor as predicted by the Zel'dovich
approximation based on the smoothed initial conditions.
In the bottom panel, the linear spin growth has been stopped at the same 
epoch for all haloes, $t_{\rm TTT} = 0.35\, t_0$.
The thick histogram corresponds to standard TTT, with a scatter of
a factor of two.
The thin histogram (which has been slightly displaced to the right) 
corresponds to $\lmod$ computed directly from
the Zel'dovich velocities at $t_{\rm TTT}$; it does not involve
the approximations made in TTT at the initial conditions 
and can be interpreted as reflecting only the scatter
due to the choice of $t_{\rm TTT}$.
}
\label{fig:ampz0}
\end{figure}

In \fig{ampz0}, we show the distribution of the ratio of 
model to true 
$L$ values at $z=0$, with the TTT growth stopped in three different ways. 
In the top panel, the TTT growth of each halo has been 
stopped at a fixed fraction of its turn-around time $t_{\rm ta}$, 
as predicted by the spherical collapse model based on the value of the
initial density fluctuation smoothed on the scale of the proto-halo
(namely, the time at which the linearly extrapolated density contrast 
is $\delta=1.06$).
By stopping the TTT growth for each halo at $t_{\rm TTT}=0.56\, t_{\rm ta}$,
the median of the ratio $\lmod/\ltrue$ becomes unity.
Alternatively, in the central panel, the TTT growth for each halo
has been stopped at a fixed fraction of the time $t_{\rm 1D}$ when 
the smoothed proto-halo is predicted by the Zel'dovich
approximation at the initial conditions to collapse along 
the first principal axis of the deformation tensor.
Requiring that the median of the corresponding amplitude ratio is equal to 
unity gives $t_{\rm TTT}= 0.39\, t_{\rm 1D}$.
In the bottom panel, the linear spin growth has been stopped at the   
same epoch for all the haloes, $t_{\rm TTT}= 0.35\, t_0$ corresponding to 
$z=1.03$, which has been determined again to bring the median value           
of $\lmod/\ltrue$ to unity.
As at $z=3$, we show the histograms for TTT and for the pure Zel'dovich
velocities at $t_{\rm TTT}$.
The corresponding percentiles are listed in Table \ref{Apct}. 
Note that, despite the different stopping times, the three probability
distributions are very similar; the 68.3 per cent confidence level 
corresponds to a factor of 2 random error. 
The histograms for TTT and for the pure Zel'dovich velocities
show a similar scatter, indicating that the choice of $t_{\rm TTT}$
is the dominant factor in determining the scatter compared to the
other approximations involved in TTT.

The impression 
from \fig{meanevo_amp} 
that the growth of spin amplitude 
really stops at some specific epoch is somewhat misleading.
When we investigate halo by halo, we find that about 38 per cent 
of our haloes actually loose angular momentum between $z=0.6$, $z=0.2$
and $z=0$ due to non-linear interactions 
(see a similar finding by Sugerman \etal 2000).
On average, these haloes loose $\sim 30$ per cent of their spin between
$z=0.6$ and $z=0$.
This is partly a result of considering in each proto-halo the same group 
of particles at all times, based on their association with a halo at $z=0$.
Angular momentum from these proto-halo particles may be transferred to 
outer shells of matter.  Mergers, tidal encounters and the presence of 
substructure all contribute to angular-momentum transfer.
This decrease of $L$ with time at late epochs indicates that in many case
the TTT growth actually continued roughly until turn-around, stopped, 
and started declining thereafter. This is one of the reasons for why 
the effective time for stopping the TTT growth should be somewhat before 
turn-around.

Some fraction of the haloes have their spin growing also at late epochs:
about 21 per cent of our spins grow between $z=0.6$, $z=0.2$ and $z=0$.
On average, these haloes increase their spin by
$\sim 60$ per cent between $z=0.6$ and $z=0$.
However, for about 57 per cent of our haloes the evolution of spin amplitude
in the non-linear regime is weak, less than 20 per cent between $z=0.6$
and $z=0$, in pleasant agreement with the concept of TTT.
Considering the whole halo population, we find that the mean 
value and the scatter of the ratio $L(z=0.6)/L(z=0)$ are $1.19$ and 
0.39 -- or $1.01$ and 0.34 for the inverse ratio.
We conclude that even though non-linear effects do not 
significantly change the average amplitude of halo spins, 
the variety of evolutionary paths result in a large scatter
about the mean TTT prediction at $z=0$.

\begin{figure}
\epsfxsize= 8 cm \epsfbox{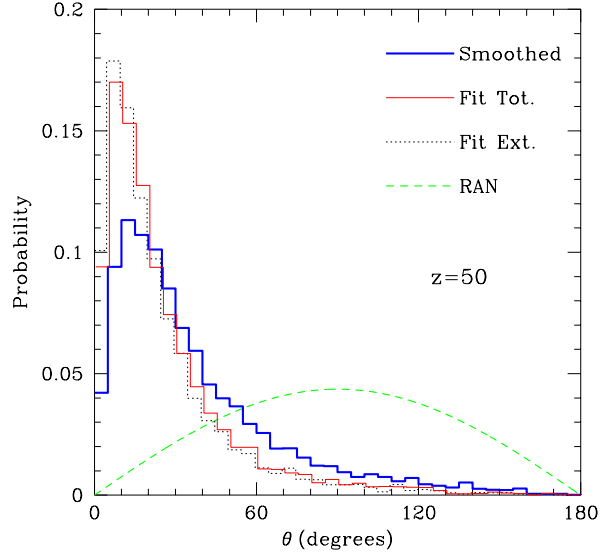}
\caption{ 
The distribution of angles between  
model and true 
spin directions in the linear regime, $z=50$.  The three histograms 
(slightly shifted horizontally to improve readability) correspond to 
the different methods of computing the shear tensor (\se{implement}). 
The dashed line is the expected distribution for random orientations.
The misalignment is due to the truncation of the Taylor expansion
of the potential. Additional error is added by the smoothing
on the proto-halo scale.
}
\label{fig:dirz50}
\end{figure}

The bottom line is that standard TTT provides a successful order-of-magnitude 
estimate for the halo angular-momentum amplitude. 
If TTT growth is stopped at turn-around, the systematic overestimate 
of the final spin is of the order of the signal. 
This systematic overestimate is removed if the TTT growth is stopped at about 
56-70 per cent of the turn-around time, or about 23-35 
per cent of the collapse time,
depending on how these times are defined. 
However, the random scatter from halo to halo is of the order of the signal 
itself. Part of this scatter is due to real non-linear evolution, reflected
in an uncertainty in the choice of the effective time at which the TTT 
growth should be stopped, and part is due to the approximations involved 
in applying TTT already at the initial conditions.
The standard truncation of the potential and smoothing of the shear tensor
is close to the best one can do in TTT.

\begin{figure}
\epsfxsize= 8 cm \epsfbox{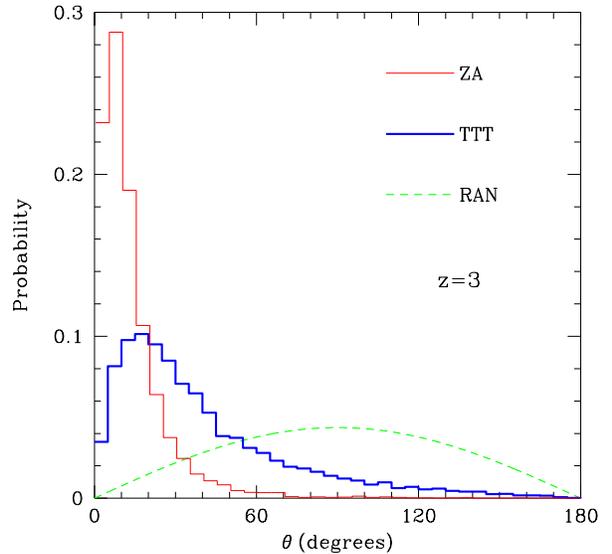}
\caption{
The distribution of the angles between 
model and true 
spin directions in the quasi-linear regime, at $z=3$.
The thick histogram is for the standard, truncated and smoothed TTT.
The thin histogram (which has been slightly displaced to the right) 
is for the angular momentum derived from the Zel'dovich velocities,
i.e., its predicted direction at any time is the true spin direction at $z=50$.
The deviation of this Zel'dovich histogram from perfect alignment 
reflects weak non-linear temporal variations in the spin directions.
The dashed curve is the distribution of angles in the case of random
orientations.
}
\label{fig:dirz3}
\end{figure}

\begin{figure}
\epsfxsize= 8 cm \epsfbox{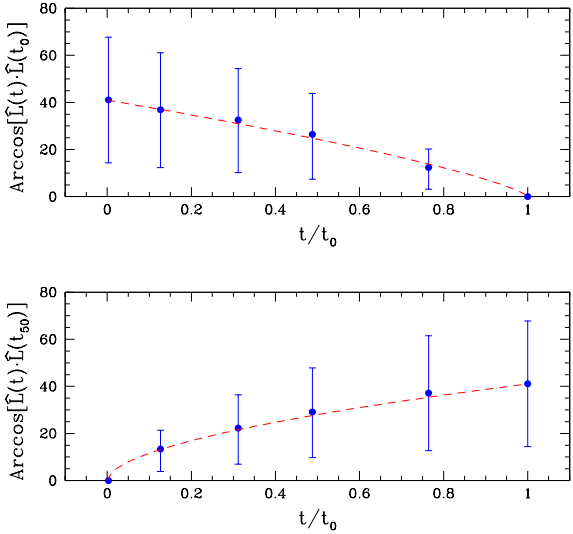}
\caption{
Evolution of true spin direction.
Shown are the average and the 68.3 per cent confidence interval over the haloes
of the angle between the angular momentum at time $t$ and its counterpart
evaluated either at the present time ($t_0$, top panel)
or at the initial conditions ($t_{50}$, bottom panel). 
Linear TTT predicts perfect alignment.
The dashed lines show empirical functional fits: 
$41^\circ (1-t/t_0)^{0.75}$ and $41^\circ 
(t/t_{0})^{0.55}$ respectively.
}
\label{fig:meanevo_dir}
\end{figure}

\begin{figure}
\epsfxsize= 8 cm \epsfbox{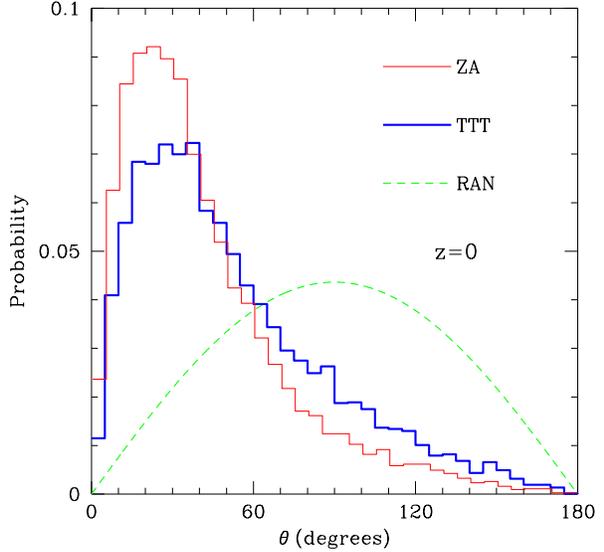}
\caption{ 
As in \fig{dirz3}, but at $z=0$.
The distribution of angles between the true spin directions
at $z=3$ and $z=0$ (not shown) practically coincides with the 
thin histogram, which measures the non-linear temporal              
variations in spin direction.
}
\label{fig:dirz0}
\end{figure}

\section{Spin Direction}
\label{sec:direction}

The fact that the TTT prediction of the spin amplitude is only good to order
of magnitude raises 
doubts about the prospects of TTT as a predictor
for the direction of the spin vector, because 
a random error of a few tens of percents 
in each component is likely to be associated with a non-negligible loss of 
information about the direction. 
We compare the predicted and true directions of the spin vector for each
halo via the distribution of the angle $\theta$ between them
over the haloes, proceeding from the linear regime at $z=50$,
to the quasi-linear epoch $z=3$, and to the present non-linear epoch $z=0$,
as we did for the spin amplitude.

In \fig{dirz50} we compare the three methods of computing the shear tensor
(\se{implement}) at the initial conditions, $z=50$. Zel'dovich velocities 
would have corresponded to a Dirac $\delta$ function at $\theta=0^\circ$,
while a random distribution of relative directions would correspond 
to the distribution shown by a dashed line. 
The average angle of misalignment in the linear regime is $37^\circ$, 
$26^\circ$, and $25^\circ$ for methods 1, 2, and 3, respectively. 
The corresponding scatter 
(defined, as before, as half the 68.3 per cent confidence interval) 
is $\sim 26^\circ, 18^\circ$, and $\sim 18^\circ$ in the three cases. 
Other percentiles of these distributions are given in Table \ref{Thetapct}. 
As in the case of the amplitude, the sources of misalignment in the
linear regime are the second-order truncation of the potential and the 
smoothing.
Considering only the haloes with 1000 particles or more, we obtain  
only slightly smaller misalignments, with mean values of
$33^\circ$, $24^\circ$, $23^\circ$ and scatter of 
$24^\circ, 17^\circ, 15^\circ$ for the three methods, respectively. 

\begin{table*}
\begin{minipage}{123mm}
\caption{Characteristics of the distributions of 
the angle between the vectors of the spins as 
predicted by the model and as measured in the simulations.
Listed are the mean and several percentiles, as well as the scatter
as measured by half the central 68.3 per cent region.
TTT=tidal-torque theory, NB=$N$-body simulations, ZA=Zel'dovich 
approximation.} 
\label{Thetapct}
\begin{tabular}
    {@{}lccccccccc}
Comparison  & $z$ 
& Fig. & {\bf Mean}
	& $2.5\%$ & $15.9\%$ & ${\mathbf 50\%}$   & $84.2\%$
  & $97.5\%$ & {\bf Scatter}\\
\\
TTT/NB  & 50 
&  7 & {\bf 36.8} & 3.9 & 11.0 &{\bf 27.3} & 63.5 & 123.6 & {\bf 26.2}\\  
TTT(2)/NB  & 50 
&  7 &{\bf  25.8} & 2.3 & 7.1 & {\bf 18.1} & 43.8 & 99.1 & {\bf 18.3} \\ 
TTT(3)/NB  & 50 
&  7 & {\bf 24.7} & 2.3 & 6.6 & {\bf 17.0} & 41.7& 95.5 & {\bf 17.6} \\
\\
ZA/NB & 3 
&  8 &{\bf 13.4} & 1.3 & 3.9 & {\bf 9.6} & 21.4& 48.6 & {\bf 8.7}\\
TTT/NB & 3 
&  8 & {\bf 40.1} & 4.1 & 12.3 & {\bf 30.3} & 69.9 & 129.0 &{\bf  28.8}\\
\\
NB/NB& 50/0
&  - & {\bf 41.1} & 5.1 & 14.4 & {\bf 33.1} & 67.7 & 122.6 & {\bf 26.6}\\
NB/NB& 3/0
&  - & {\bf 36.9} & 4.5 & 12.3 & {\bf 29.1} & 61.1 & 116.3 & {\bf 24.4} \\
NB/NB& 1.2/0
&  - & {\bf 32.5} & 3.5 & 10.2 & {\bf 24.3} & 54.4 & 111.8 & {\bf 22.1} \\
NB/NB& 0.6/0
&  - & {\bf 26.4} & 2.6 & 7.4 & {\bf 18.7} & 43.8 & 100.3 & {\bf 18.2} \\
NB/NB& 0.2/0
&  - & {\bf 12.4} & 1.1 & 3.1 & {\bf 7.9} & 20.2 & 52.1 & {\bf 8.5} \\
NB/NB& 0/3
&  - & {\bf 36.9} & 4.5 & 12.3 & {\bf 29.1} & 61.1 & 116.3 & {\bf 24.4}\\
\\
ZA/NB &0
&  10 & {\bf 41.1} & 5.1 & 14.4 & {\bf 33.1} & 67.7 & 122.6 & {\bf 26.6} \\
\\
TTT/NB & 0 
&  10 & {\bf 52.2} & 7.1 & 18.7 & {\bf 43.3} & 89.0 & 139.3 & {\bf 35.1}\\
\end{tabular}
\end{minipage}
\end{table*}

We find that considering the third-order term in the expansion of the 
gravitational potential within the proto-halo region (method 1, 
not shown in the figure) results in a very small improvement: 
the average is reduced from $37^\circ$ to $34^\circ$,
and the scatter from $26^\circ$ to $24^\circ$.
Considering the haloes for which standard TTT does a poor job and
mispredicts the spin direction by more than $60^\circ$, we find that 
by adding the third-order term one improves the prediction in 70 per cent 
of the cases. For the remaining haloes, where TTT is doing a relatively
reasonable job, the third-order term does not lead to a general improvement
and it improves the prediction in only 54 per cent of the cases. 

It is interesting to check, already in the linear regime,
whether a good prediction for the spin direction
is associated with a good prediction for the spin amplitude.
For this purpose, we split the halo sample into a ``good" group and a
``bad" group according to
whether the spin direction predicted by standard TTT and the true 
spin direction differ by less or more than $30^\circ$,
containing 54 and 46 per cent of the haloes respectively.
For the ``good" group, the average and 68.3 per cent confidence interval of 
$\lmod/\ltrue$ are $0.91$ and $(0.58,1.18)$, and for the `bad' group they are
$1.00$ and $(0.44,1.44)$. 
We learn that erroneous predictions for the spin direction correspond
to a large scatter in the amplitude, while more accurate predictions
for the direction tend to systematically underestimate the amplitude.
We find that for 31 per cent of our haloes, the TTT estimates lie both
within $30^\circ$ of the true direction and differ from the correct
amplitude by less than 30 per cent.
 
The goodness of TTT predictions correlates with the halo 
mean specific angular momentum.
For instance, considering the 10 per cent of the haloes with the
lowest specific angular momentum, we find that standard TTT 
overestimates the spin amplitude by an average (median) factor of 1.6
(1.22) with a scatter of $\pm 0.92$, and mispredicts
the spin direction by $67^\circ$ ($60^\circ$) $\pm 40^\circ$.
On the other hand, for the 10 per cent of the haloes with the highest
specific angular momentum, we find that the mean (median) and scatter
are 0.87 (0.83) $\pm 0.24$ for the amplitude ratio
and $18^\circ$ ($13^\circ$) $\pm 11^\circ$ for the angle.
These statistics vary continuously as a function of the
halo specific angular momentum. 
A similar behaviour is found using methods 2 and 3 of \se{implement},
but in these cases TTT does not introduce a systematic error for the 
haloes with the highest specific angular momentum. 
This behaviour seems to suggest that TTT introduces a typical angular-momentum
error per particle,
and good predictions can be only obtained when the halo mean specific angular 
momentum is larger than this typical uncertainty.

In \fig{dirz3} we test how well the spin maintains the constant direction
implied by TTT during the quasi-linear evolution, until $z=3$. 
The spin derived from the Zel'dovich velocities at $z=3$,
which is of course parallel to the spin at the initial conditions,
has an average misalignment with the true spin of only $13^\circ$
with a comparable scatter. 
This small misalignment implies that the spin tends to keep a pretty fixed
direction during quasi-linear evolution.
When this small temporal variation of the angle is convolved with 
the truncation and smoothing involved in applying TTT to the initial
conditions, the misalignment becomes more significantly larger, 
with an average of $40^\circ$ and a scatter of $\sim 30^\circ$. 
The mean evolution and scatter of the true halo spin direction 
all the way to $z=0$ is summarized in
\fig{meanevo_dir}, either relative to the direction at $z=0$ or
relative to the direction at $z=50$. Empirical functional fits are 
$41^\circ (1-t/t_0)^{0.75}$ and $41^\circ (t/t_{0})^{0.55}$ respectively.
We see a significant true misalignment developing due to non-linear effects.

\fig{dirz0} addresses the full non-linear variation of spin direction
down to $z=0$.
When the model spin is computed from the particles following the Zel'dovich 
approximation, we obtain the true total misalignments due to non-linear 
effects, 
as in \fig{meanevo_dir}, 
with an average and scatter of $41^\circ$ and $27^\circ$.
This true misalignment occurs mostly between $z=3$ and $z=0$:
this part of the misalignment (not shown in the figure) has an average and
scatter of $37^\circ$ and $24^\circ$, almost as large as the total true
misalignment.

Finally,
the full TTT prediction, with the additional truncation and smoothing, 
yields a mean misalignment with the actual spin at $z=0$ of 
$52^\circ$ with a scatter of $\sim 35^\circ$. 
We see that most of the misalignment is 
due to true non-linear evolution of spin direction. 
This implies that no matter how exactly TTT is applied, it is
doomed to be a limited predictor of spin direction in the non-linear regime.

The bottom line is that non-linear evolution causes significant variations
in spin direction, which limit the quality of linear TTT as an accurate
predictor of spin direction. Still, there is a significant correlation between
the direction predicted by TTT and the final spin direction.

\section{Spin--Spin Correlations}
\label{sec:L-L}

TTT has been used to predict spatial spin correlations
as indicators of intrinsic spatial alignments of galaxies 
in the context of interpreting weak gravitational lensing signals 
(Catelan \etal 2001; Crittenden \etal 2001).
Our results of the previous section already indicate that TTT is a poor 
predictor of galaxy spin direction.
In order to address directly the usefulness of TTT as a predictor of
spatial spin correlations, we compute in this section two-point spin 
statistics.

A summary of the relevance to lensing studies is as follows.
A shear signal, interpreted as a result of gravitational lensing
due to the large-scale distribution of matter, has been detected
by several groups (Bacon \etal 2000; Kaiser \etal 2000;
van Waerbeke \etal 2000, 2001; Wittman \etal 2000).
Lensing shear maps can be used to directly reconstruct the mass distribution,
free of galaxy biasing. However, a significant uncertainty arises
due to possible intrinsic coherent alignment of galaxy shapes.  
Indeed, Brown \etal (2000) (see also Pen \etal 2000) claim to have
detected a non-vanishing correlation between galaxy ellipticities 
in the local universe, over a range of angular separations between 
1 and 100 arc-minutes.  Since the galaxy samples used for these studies 
are relatively nearby, a lensing interpretation is highly improbable, 
so this signal is likely to reflect intrinsic galaxy alignments.

The simplest theoretical approach is to link galaxy shapes to halo spins
by assuming that galaxy discs are perpendicular to the spin vector of 
their host haloes.  The spins are expected to be correlated because
the tidal field, that plays a key role in generating 
the angular momentum,
is expected to have a non-negligible coherency over large scales 
(Catelan \& Porciani 2001).
Alternative models assume  that the light profile of
elliptical galaxies approximate the shape of their hosting haloes
(Croft \& Metzler 2000; Heavens \etal 2000). 

We introduce two statistics for measuring two-point spin correlations.
The two-point correlation function of spin directions is simply,
\begin{equation}
\label{eq:sc1}
\eta(r)=\langle \hat{L}(\bfx)\cdot\hat{L}(\bfx +\bfr) \rangle\;.
\end{equation}
The function
\begin{equation}
\label{eq:sc2}  
\eta_2(r)=\langle [\hat{L}(\bfx)\cdot\hat{L}(\bfx +\bfr)]^2
\rangle-\f{1}{3}
\end{equation}
measures the correlation of spin axes independent of the sense of rotation,
and is therefore more directly relevant to the intrinsic correlations
of shape orientations that mimic lensing signal.

In \fig{spineta} we compare the functions $\eta(r)$ and $\eta_2(r)$,
at different epochs of the simulation, with the prediction of TTT as 
evaluated from the initial density and velocity fields.\footnote{The 
correlation functions presented by
Heavens \etal (2000) and by Croft \& Metzler (2000) seem to be
less noisy than ours, but this is an illusion caused by the fact that they
allowed haloes smaller than we do, down to 10 particles, while the resolution
of their simulations is similar to ours.
By including more haloes in the sample, one may be tempted to assume that
the statistical errors are reduced, but the error budget must include
the very large errors associated with measuring spin and shape for each
halo based on such a small number of particles (see Bullock \etal 2001).}
We see in the simulation that at high redshift there is a positive
spin--spin correlation at separations of a few comoving megaparsecs, but 
the correlation signal on these scales 
is much weaker at $z=0$. 
This is mainly due to non-linear effects. 
On the other hand, the TTT approximation 
maintains the same correlation function at all times,
because it keeps the spin directions fixed, by construction. 
At $z=50$, TTT matches well the simulation data,
and since it always builds upon the initial conditions, it maintains
the same 
correlation signal at late times as well. The linear TTT approximation
naturally fails to include the non-linear effects that weaken the 
correlations 
on scales $\geq 1\hmpc$, 
and it thus overestimates the correlation 
at late times.\footnote{
We also find that adding the third term in the Taylor expansion of the 
potential -- see \equ{LZel3} -- leaves the prediction for the correlation
functions almost unaffected (not shown in the figure).}

The weakening of the correlation signal on the scales shown can be
attributed to two non-linear effects. First, the members of galaxy pairs tend
to get closer due to the evolving clustering. Second, the spin directions
themselves evolve away from the TTT predictions.  As a result of the first
effect, the linear correlation signal simply shifts to smaller scales.
The interpretation of the second could be that non-linear dynamics erases the 
linear spin-spin correlations on all scales, while non-linear halo-halo 
interactions build up a new correlation signal on small scales 
(note, for example, that for $r\lsim 2 \hmpc$, $\eta$ is positive at $z=50$, 
while it is not positive at $z=0$). 
We distinguish between these effects by also showing in \fig{spineta} 
the spin-spin correlation functions as computed using the TTT predictions
for the halo spins but located at the actual halo positions at $z=0$.
Even though this model does a better job than linear TTT,
it still over-predicts the correlations measured in the simulations at $z=0$.
This demonstrates that non-linear halo-halo interactions and 
angular-momentum exchanges with additional infalling material play an 
important role in determining spatial spin correlations on small scales.
Using a simulation of higher resolution,
one can indeed detect significant spin correlations at $z=0$ on scales
below $0.4\hmpc$ (Dekel \etal 2000).

Since present-day galactic discs have been probably assembled 
sometime around $z=1$, 
we also compute spin-correlations for haloes selected at that epoch.
In \fig{spincorz1}, we show the
results for $\sim 5400$ haloes containing 
100 particles or more (left panel), and for $\sim 68000$ haloes
containing at least 10 particles (right panel). 
Linear correlations are significantly 
weakened by non-linear effects 
already at $z=1$. 

It is worth mentioning that
the amplitude of the correlation function $\eta_2(r)$ for haloes
selected at $0\leq z \leq 1$ in our simulation
is in good agreement with the observed alignment of galaxy shapes 
in the local galaxy population (e.g. see Figure 1 in Pen \etal 2000).
This indicates that spin correlations between dark matter haloes
might be related to the correlations of galaxy shapes. 

In summary: our results show that linear TTT over-predicts 
spin-correlations for non-linear dark matter haloes
by a factor of a few. Therefore, caution should be applied when using
standard TTT to predict the effects of intrinsic alignments in weak
lensing studies.

\section{Another Model for Spin-Spin Correlations}
\label{sec:ss2}

An alternative model for spin-spin correlations have been proposed
by Lee \& Pen (2000, LP) and Pen \etal (2001). 
It is also based on linear perturbation theory, but it tries to bypass
the difficulties associated with determining the inertia tensor 
of protohaloes by assuming that the spin directions are statistically
correlated with the local tidal field in a specific way. 
The basic Ansatz is 
\begin{equation}
\langle \hat{L}_i \hat{L}_j | \hat{T}_{ij}\rangle=
\f{1+a}{3}\delta_{ij}-a\hat{T}_{ik}\hat{T}_{kj}\;,
\label{eq:leepen}
\end{equation}
where $\hat{T}_{ij}=T_{ij}/[T_{lk}T_{kl}]^{1/2}$ is the trace-free
unit shear tensor, and $a$ is a correlation parameter, whose value is
to be determined.
The average is over all possible inertia tensors that
are compatible with a given realization of the tidal field.

One then assumes that the inertia tensors of protohaloes at different
locations are statistically independent (which is questionable,
since the inertia tensors are found to be strongly correlated with
the tidal field; see Paper II). 
This allows one to perform the ensemble average
in two steps, first averaging over the inertia tensors for a
given tidal field, and then over the realizations of $T_{ij}$.
This leads to
\begin{equation}
\eta_2(r) \propto \langle \hat{T}_{ik}(\bfx) \hat{T}_{kj}(\bfx) 
\hat{T}_{il}(\bfx') \hat{T}_{lj}(\bfx') \rangle, \quad  |\bfx'-\bfx|=r.
\end{equation}
While $\hat{T}_{ik}$ is non-Gaussian, $T_{ik}$ is assumed to be Gaussian,
so one can proceed with the calculation if one is willing to assume
further that
\begin{eqnarray}
\label{eq:tthat}
\langle \hat{T}_{ik}(\bfx) \hat{T}_{kj}(\bfx) 
\hat{T}_{il}(\bfx') \hat{T}_{lj}(\bfx') \rangle
&\simeq& \\
&\simeq &
\f{
\langle {T}_{ik}(\bfx) {T}_{kj}(\bfx) 
{T}_{il}(\bfx') {T}_{lj}(\bfx') \rangle}{\langle T_{lk}T_{kl} \rangle^2}.
\nonumber
\end{eqnarray}
Applying Wick's theorem, LP obtained this way the approximation
\begin{equation}
\eta_2(r)\simeq \f{a^2}{6} \left[\f{\xi_M(r)}{\sigma_M^2}\right]^2, 
\label{eq:lp_eta2}
\end{equation}
where $\xi_M(r)$ and $\sigma_M^2=\xi_M(0)$ are respectively the two-point
correlation function and the variance of the mass overdensity field
smoothed on the typical proto-halo scale.
By including all the terms of the same order as $\xi_M^2$,  
Catelan \& Porciani (2001; equation 20) provided the more complete
expression:
\begin{eqnarray}
\eta_2(r)&\simeq& \f{9 a^2}{4 \sigma_M^4} \left(
\f{4}{9} \xi_M^2+\f{8}{9}J_{3,M} \xi_M-\f{8}{5}J_{5,M} \xi_M+\f{14}{9}
J_{3,M}^2 -\right.\nonumber \\
&-&4 \left.J_{3,M} J_{5,M}+\f{14}{5} J_{5,M}^2 
\right)\;, 
\label{eq:catpor}
\end{eqnarray}
where $J_{n,M}(r)=n \,r^{-n} \int_0^r dq\, q^{n-1} \xi_M(q)$.
For scale-free power spectra, $\eta_2$ is indeed proportional to $\xi_M^2$,
as in \equ{lp_eta2}.

\begin{figure*}
\centerline{ 
\epsfxsize= 8 cm \epsfbox{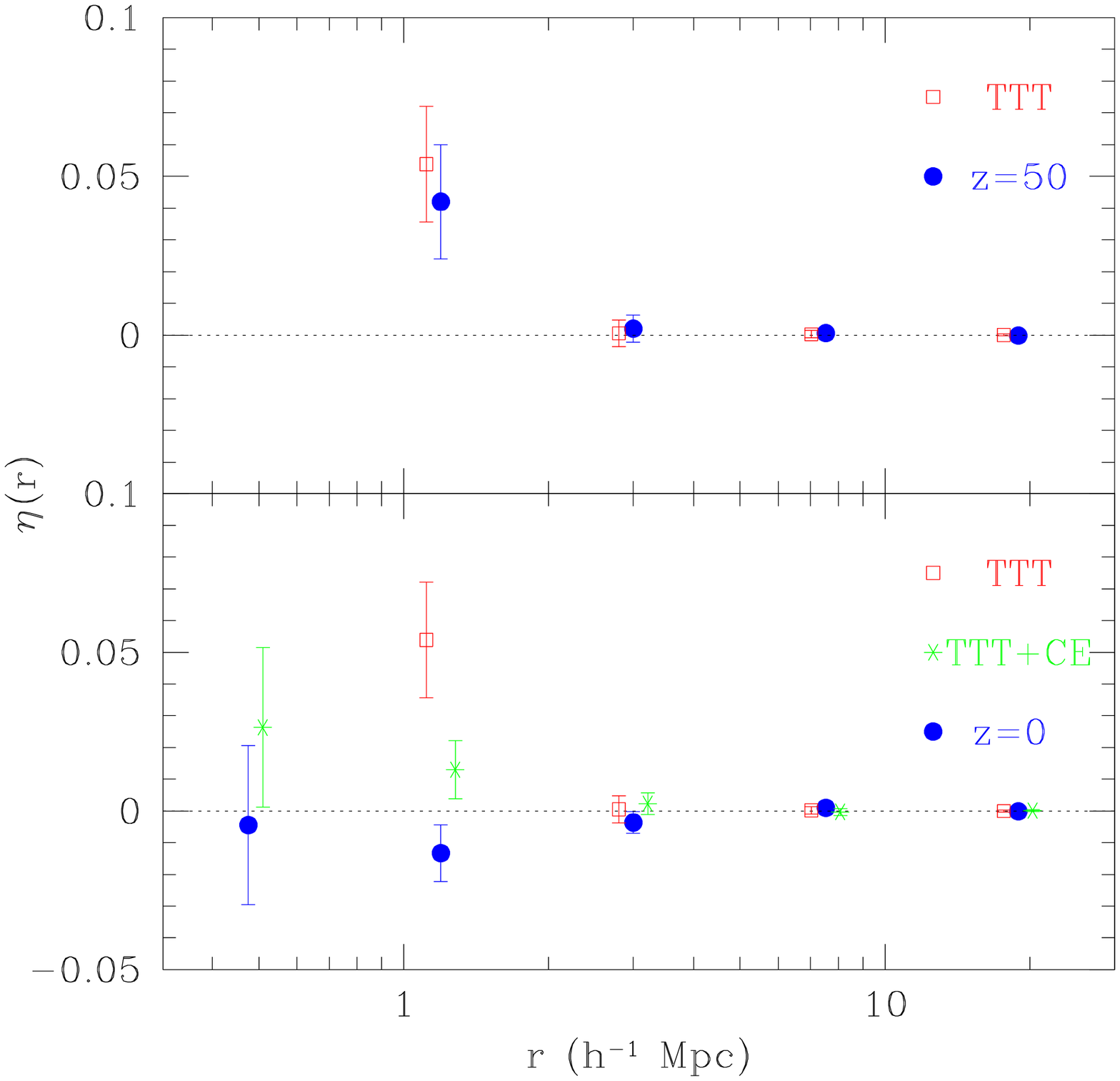} 
\epsfxsize=8 cm \epsfbox{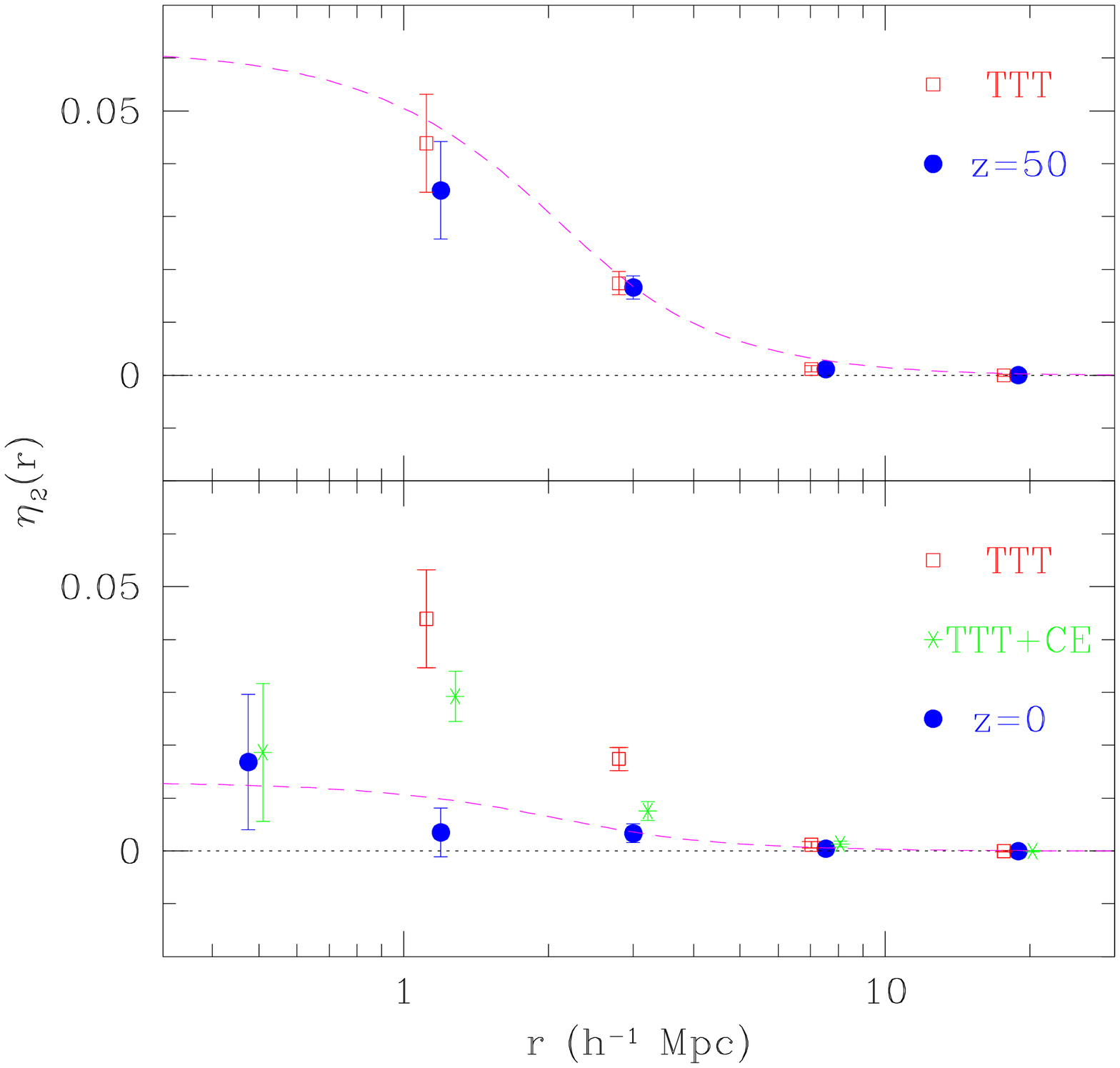}}
\caption{ Spin--spin correlation functions, as defined in \equ{sc1} (left)
and \equ{sc2} (right).
The time-independent TTT predictions (open squares) 
are compared with the true simulation data 
(filled circles) at $z=50$ (top) and $z=0$ (bottom).
The effect of allowing for clustering evolution (CE) by associating the TTT
predictions with the halo positions at $z=0$ is also shown 
(stars). 
The TTT predictions were artificially displaced by $\pm0.03$ in $\log (r)$
for clarity.
Error-bars denote one-standard-deviation statistical uncertainties.
The spatial separation $r$ is in comoving units} 
\label{fig:spineta}
\end{figure*}

\begin{figure*}
\centerline{ 
\epsfxsize= 8 cm \epsfbox{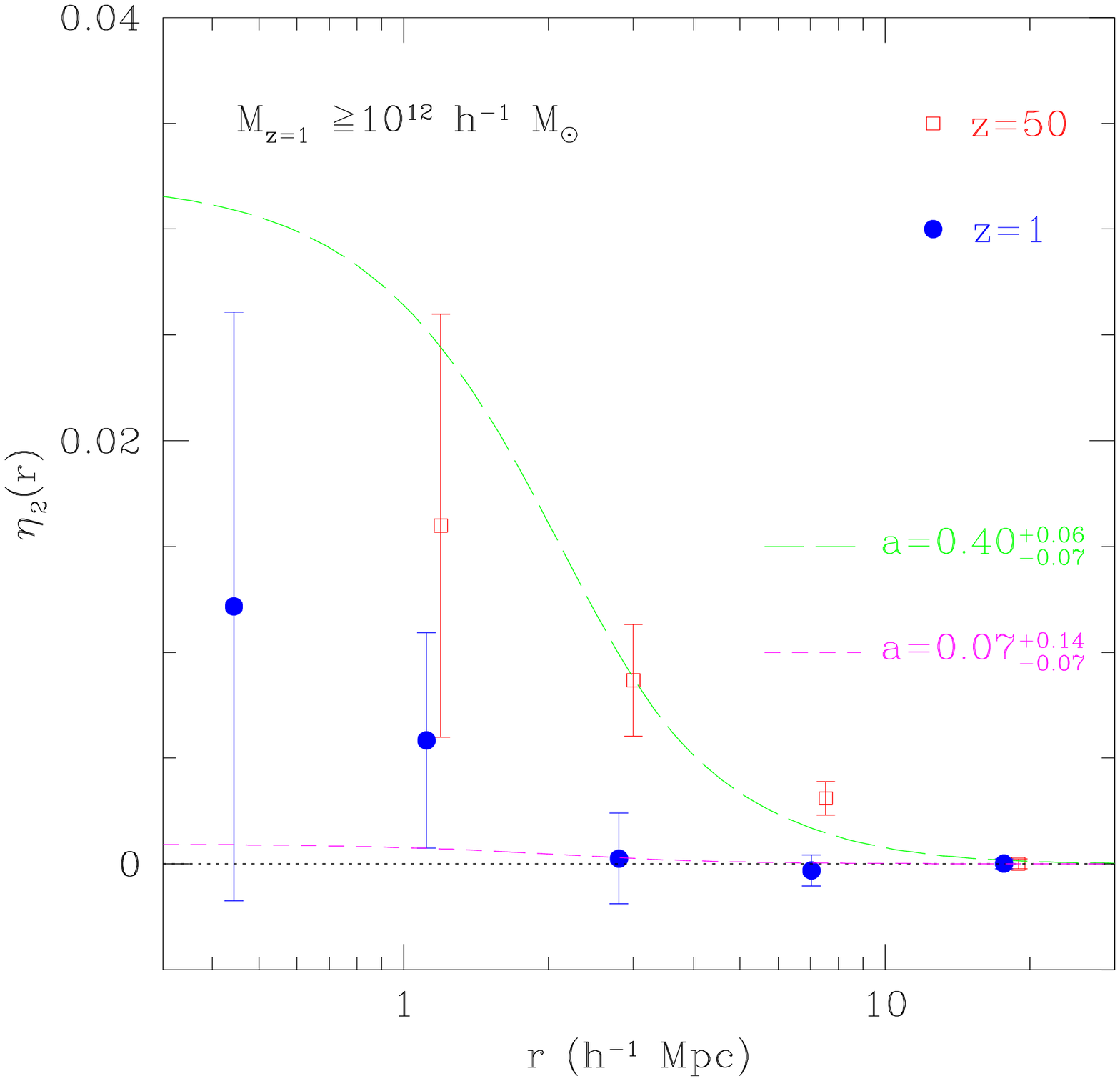} 
\epsfxsize=8 cm \epsfbox{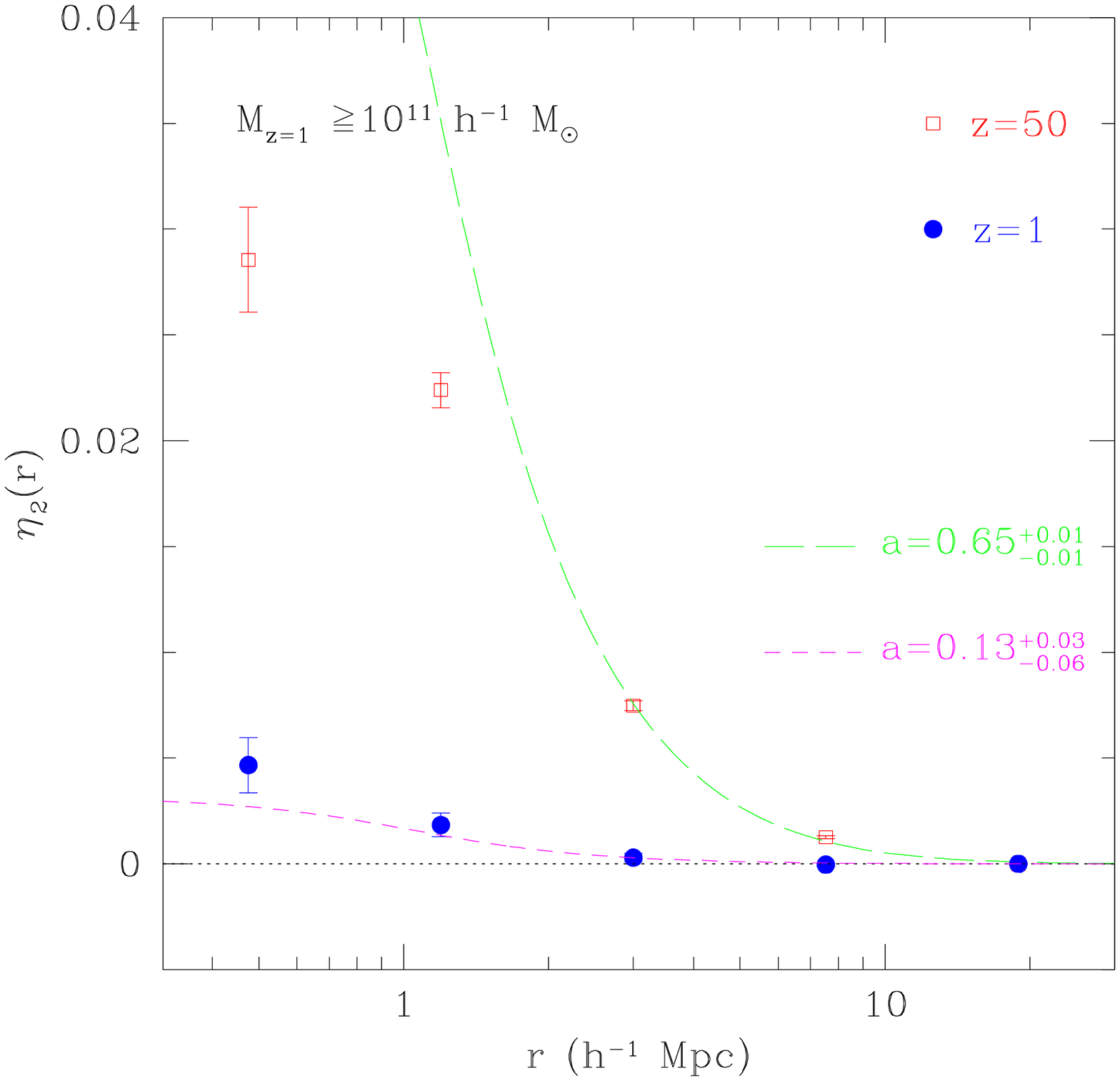}}
\caption{ Spin--spin correlation functions, as defined in \equ{sc2}
for haloes selected at $z=1$ and which contain at least 100 (left)
or 10 (right) particles.
The simulation data at $z=50$ (open squares) 
are compared with those at $z=1$ (filled circles).
The open squares in the left panel were artificially displaced by 
$-0.03$ in $\log (r)$ for clarity.
Error-bars 
refer to the sampling errors only, and ignore the error associated with
the individual spin measurements.
The dashed line shows the approximation for $\eta_2(r)$ given
in \equ{catpor}. The value of the parameter $a$ is determined
by requiring analytical and numerical correlations to coincide
at $r\simeq 3 \hmpc$.
The spatial separation $r$ is in comoving units} 
\label{fig:spincorz1}
\end{figure*}

Spin-spin correlations as predicted by \equ{catpor} are compared to the 
simulation results in \fig{spineta} and \fig{spincorz1}.
We see that there is a reasonable qualitative agreement, but not in
quantitative detail.
The smoothing length adopted to compute the analytic prediction
corresponds to the mass of the smallest (and most abundant) 
haloes in each figure. 
Note that, because of the $\sigma_M^4$ term at the denominator, the analytic
result for $\eta_2$ (with a fixed $a$) 
depends strongly on $M$ for separations much larger than
the smoothing length.
The value of the parameter $a$ is fixed by matching the
prediction and the numerical result at $r\simeq 3 \hmpc$. 
For the haloes selected at $z=0$, it is
$a=0.56^{+0.03}_{-0.04}$ at $z=50$, and 
$a=0.26^{+0.05}_{-0.08}$ at $z=0$.
However, a direct estimation of $a$ in \equ{leepen}, using the
$L-T$ correlations found in the same simulation in Paper II,
yields different results. For example, considering haloes selected at $z=0$,
we find $a=0.28 \pm 0.01$ at $z=50$, and $a=0.07\pm 0.01$ at $z=0$,
namely, a difference by a factor of two to four.
This difference seems to indicate that either \equ{leepen}
or some of the simplifying assumptions leading to \equ{catpor}
are not good approximations.

It is also worth noticing that, for a given spin correlation function,
the value of $a$ in \equ{catpor} must depend on the scale over
which the tidal field is smoothed. We find that this dependence is likely to be
spurious, because when we evaluate $a$ directly from \equ{leepen} using 
two alternative smoothing kernels (either corresponding to the halo mass 
or 8 times bigger) we see no significant scale dependence (Paper II, \S 7).
The dependence on the smoothing radius implied by \equ{catpor} probably 
originates from the questionable assumption made in \equ{tthat}.
This ambiguity regarding the scale dependence may affect the interpretation 
of the observed tidal field, which can be measured reliably only on scales 
much larger than the galactic scales where $a$ is defined.  For example, 
Lee \& Pen (2001) measure $a \sim {\rm a\ few} \times 10^{-2}$ from the 
PSCz data smoothed by Wiener Filter on large scales, and then obtain
$a\sim 0.2$ by extrapolation to galactic scales using the ambiguous
scale dependence implied by \equ{lp_eta2}. 

By construction, \equ{catpor} does not account for non-linear
effects, which affect both the spin direction and the clustering growth,
but we see that its accuracy is limited even in the linear regime.
The fact that \equ{catpor} with $a\sim 0.1-0.2$
provides an approximate description of spin correlations 
in our simulation at $0\leq z\leq 1$ and in observational data
should therefore be attributed to a lucky coincidence.  

\section{Conclusion}
\label{sec:conc}

We have evaluated the performance of linear tidal-torque theory in predicting
the spin of galactic haloes using $\sim 7300$ well-resolved virialized 
dark-matter clumps extracted from a cosmological $N$-body simulation.
We defined haloes in today's density field using the standard 
friends-of-friends algorithm, 
but it would be useful for future work to check
robustness to alternative halo finders as well as to different
cosmological scenarios.

We found that, for a given proto-halo at the initial conditions, 
TTT provides a successful order-of-magnitude estimate of the final halo 
spin amplitude. The TTT prediction matches on average the spin amplitude
of today's virialized haloes if linear TTT growth is assumed until about 
$t_0/3$, 
or about 5/9 of the turn-around time for each halo. 
The random error, from halo to halo, 
is about a factor of two. 
This makes TTT useful for studying certain aspects of galaxy formation, 
such as the origin of a universal spin profile in haloes (Bullock \etal 2001b; 
Dekel \etal 2001), but only at the level of average properties.

Non-linear evolution causes significant variations in spin direction,
which limit the accuracy of the TTT predictions to a mean error of
$\sim 50^\circ$. Furthermore, spatial correlations of spins
on scales $\geq 1\hmpc$ are 
strongly weakened by non-linear effects. 
This limits the usefulness of TTT in predicting intrinsic galaxy alignments
in the context of weak gravitational lensing (Catelan \etal 2001; Crittenden
\etal 2001).  
This situation may improve if
the orientations of today's discs were determined by the halo spins at
a very high redshift, which are better described by TTT. 
On the other hand, we know this only for the haloes selected at $z=0$, and not
necessarily for the subhaloes which host disc formation at higher redshifts.
A detailed study of this effect
would require a high-resolution simulation in which a detailed galaxy 
formation scheme is incorporated.
A preliminary analysis performed by using dark matter haloes 
selected at $z=1$ showed that, also
in this case, non-linear effects decrease linear spin-spin correlations
by a factor of a few. 

For practitioners of TTT,
we found that the standard approximations made in TTT, such as the second 
order expansion of the Zel'dovich potential, and the smoothing of 
the shear tensor on the proto-halo scale, are hard to improve upon.
The inaccuracies in the TTT predictions are dominated by real non-linear 
effects rather than the above approximations.

In order to deepen our understanding of how TTT actually works,
we investigate in paper II the cross-talk between the proto-halo inertia tensor
and the external shear tensor. 
We find to our surprise that they are strongly correlated, in the sense that
their minor, major and medium principal axes tend to be aligned, in this order.
This means that the angular momentum, which plays such a crucial role
in the formation of disc galaxies, is only a residual which arises 
from the little, $\sim 10$ per cent  
deviations from perfect alignment of $T$ and $I$.  
We find there that the $T-I$ correlation induces a weak tendency of the 
proto-halo spin
to be perpendicular to the major axis of $T$, but non-linear changes in
spin direction erase almost 
any memory of the initial shear tensor, and therefore
observed spin directions cannot serve as very useful indicators
for the initial shear tensor (cf. Lee \& Pen 2000). 

On the other hand, the strong $T-I$ correlation investigated in Paper II 
provides a promising hint for how to solve a long-standing problem
in galaxy formation theory, of identifying the boundaries of proto-haloes 
in cosmological initial conditions 
(Porciani, Dekel \& Hoffman, in preparation). 

In our studies of the cross-talk between the different components of TTT,
we have also realized another surprise that leads to a revision in
the standard scaling relation of TTT (While 1984).
We find in Paper III (Porciani \& Dekel in preparation) that the off-diagonal, 
tidal terms of the deformation tensor, which drive the torque, are 
weakly 
correlated with the diagonal terms, which determine the overdensity at the 
proto-halo centre. The latter enters the TTT scaling relation via the expected
collapse time of the proto-halo, and the 
weak
correlation with the torque
leads to a modification in the scaling relation.
The revised scaling relation can be applied shell by shell, together with
an extended Press-Schechter recipe for merger history,
in order to explain the origin of the universal angular-momentum profile
of haloes (Bullock \etal 2001b; Dekel \etal 2001). It can also be very
useful in incorporating spin in semi-analytic models of galaxy formation
(Maller, Dekel \& Somerville 2002). 

\section*{Acknowledgments}

This research has been partly supported by the Israel Science Foundation
grant grants 546/98 (AD) and 103/98 (YH),  
and by the US-Israel Binational Science Foundation grant
98-00217 (AD).  CP acknowledges the support of a Golda Meir fellowship at HU 
and of the EC RTN network `The Physics of the Intergalactic Medium' at the 
IoA.
We thank our GIF collaborators, especially H.  Mathis, A.
Jenkins and S.D.M. White, for help with the GIF simulations.


\bsp
\end{document}